\documentclass[a4paper,fleqn,usenatbib]{mnras}
\usepackage{newtxtext,newtxmath}
\usepackage[T1]{fontenc}
\usepackage{ae,aecompl}

\usepackage{graphicx}	% Including figure files
\usepackage{amsmath}	% Advanced maths commands
\usepackage{amssymb}	% Extra maths symbols

\graphicspath{{figures/}}
\usepackage{enumitem}
\title[Passage of asteroid bodies]{On the possibility of through passage of asteroid bodies across the Earth's atmosphere}

\author[D.E. Khrennikov et al.]{
Daniil E. Khrennikov,$^{1}$
Andrei K. Titov,$^{2}$
Alexander E. Ershov,$^{1,3}$
Vladimir I. Pariev$^{4}$\thanks{E-mail: vpariev@td.lpi.ru}
\newauthor and Sergei V. Karpov$^{1,5,6}$\thanks{E-mail: karpov@iph.krasn.ru}
\\
$^{1}$Siberian Federal University,   Svobodny Av. 79/10, Krasnoyarsk,  660041, Russia\\
$^{2}$Moscow Institute of Physics and Technology, Institusky Per. 9, Dolgoprudny 141700, Russia\\
$^{3}$Institute of Computational Modeling SB RAS,  Akademgorodok 50/44, Krasnoyarsk, 660036, Russia\\
$^{4}$P.~N. Lebedev Physical Institute, Leninsky Prosp. 53, Moscow 119991, Russia \\
$^{5}$L.~V. Kirensky Institute of Physics, Federal Research Center KSC SB RAS,  Akademgorodok 50/38, Krasnoyarsk, 660036, Russia\\
$^{6}$Siberian State University of Science and Technology, Krasnoyarsky Rabochy Av. 31, Krasnoyarsk 660014, Russia
}

\date{Accepted 2020 January 15. Received 2020 January 12; in original form 2019 June 28}

\pubyear{2020}

\begin{document}
\label{firstpage}
\pagerange{\pageref{firstpage}--\pageref{lastpage}}
\maketitle

\begin{abstract}
We have studied the conditions of through passage of asteroids with diameters  200, 100 and 50~m, consisting of three types of materials~--- iron, stone and water ice across the Earth's atmosphere with the minimum trajectory altitude 10--15~km. The conditions of this passage with subsequent exit into outer space with the preservation of a substantial fraction of the initial mass have been found. The results obtained support our 
idea explaining one of the long-standing problems of astronomy~--- the Tunguska phenomenon which has not received reasonable and comprehensive interpretations to date. We argue that Tunguska event was caused by an iron asteroid body, which passed through the Earth's atmosphere and continued to the near-solar orbit. 
\end{abstract}

\begin{keywords}
meteorites, meteors, meteoroids -- minor planets, asteroids: general
\end{keywords}

%%%%%%%%%%%%%%%%%%%%%%%%%%%%%%%%%%%%%%%%%%%%%%%%%%

%%%%%%%%%%%%%%%%% BODY OF PAPER %%%%%%%%%%%%%%%%%%

\section{Introduction}

The problem of the motion in the Earth's atmosphere of a large space body (SB), capable of falling onto the surface of the planet in the form of meteorites, is now of great interest. Equally urgent concern is the study of the conditions for the passage of such bodies through the upper atmosphere, even without collision with the Earth's surface, since the shock waves produced by this passage have a colossal destructive effect~\citep{Stulov1995,Bronshten1981,Martin1966,Loh1963,Hawkins1964,Andruschenko2013,Nemchinov1999,Robertson2019,
Morrison2019,TomGehrels1994}.

Large SBs (1--10~km in size and larger) that carry the potential danger of collision with the Earth are detected by ordinary astronomical observations. The bodies of intermediate dimensions began to be registered relatively recently. Observation of such bodies and the interpretation of observational data make it possible to determine the probability of their collision with the Earth, their properties, and the characteristic features of passage through the atmosphere, as well as the consequences of fall. The clarification of these questions will enable to assess more accurately the degree of asteroid hazard. 

One of the fundamental problems of meteor physics is the determination of the pre-atmospheric mass of SBs, since the intensity of the meteor phenomenon is determined by the kinetic energy of the body when entering atmosphere of the planet. It is known that the velocity of the bodies belonging to the Solar system at the entrance to the Earth's atmosphere should be inside a relatively narrow range 11.2~km/s$<V_{sn}<72.8$~km/s~\citep{Bronshten1981}, so that the variance of the contribution of the velocity squared factor to the kinetic energy does not exceed 50 times. At the same time, the mass of a meteor body can vary in a much wider range: from fractions of a gram (micrometeor) to tens of millions of tons or more (the Tunguska space body), that is, by 13--15 orders of magnitude.

The goal of this paper is to evaluate the effect on the trajectory of the SB of its passage through dense layers of the atmosphere, taking into account the acting forces, the initial velocity, the mass and its variation during the flight, to determine the conditions for possible passage of a large SB through the atmosphere with a minimum loss of mass without collision with the Earth's surface. The obtained results are compared with observational data on the Tunguska space body with an estimated altitude of the maximum energy release about 10--15~km to receive evidence in favour of a new explanation of the Tunguska phenomenon which attributes the absence of meteoritic material on the Earth's surface near the epicentre to the through passage of the SB across the atmosphere with small loss of velocity. 

\section{Physical model}

\begin{figure}
\centering
	\includegraphics[width=5.5cm]{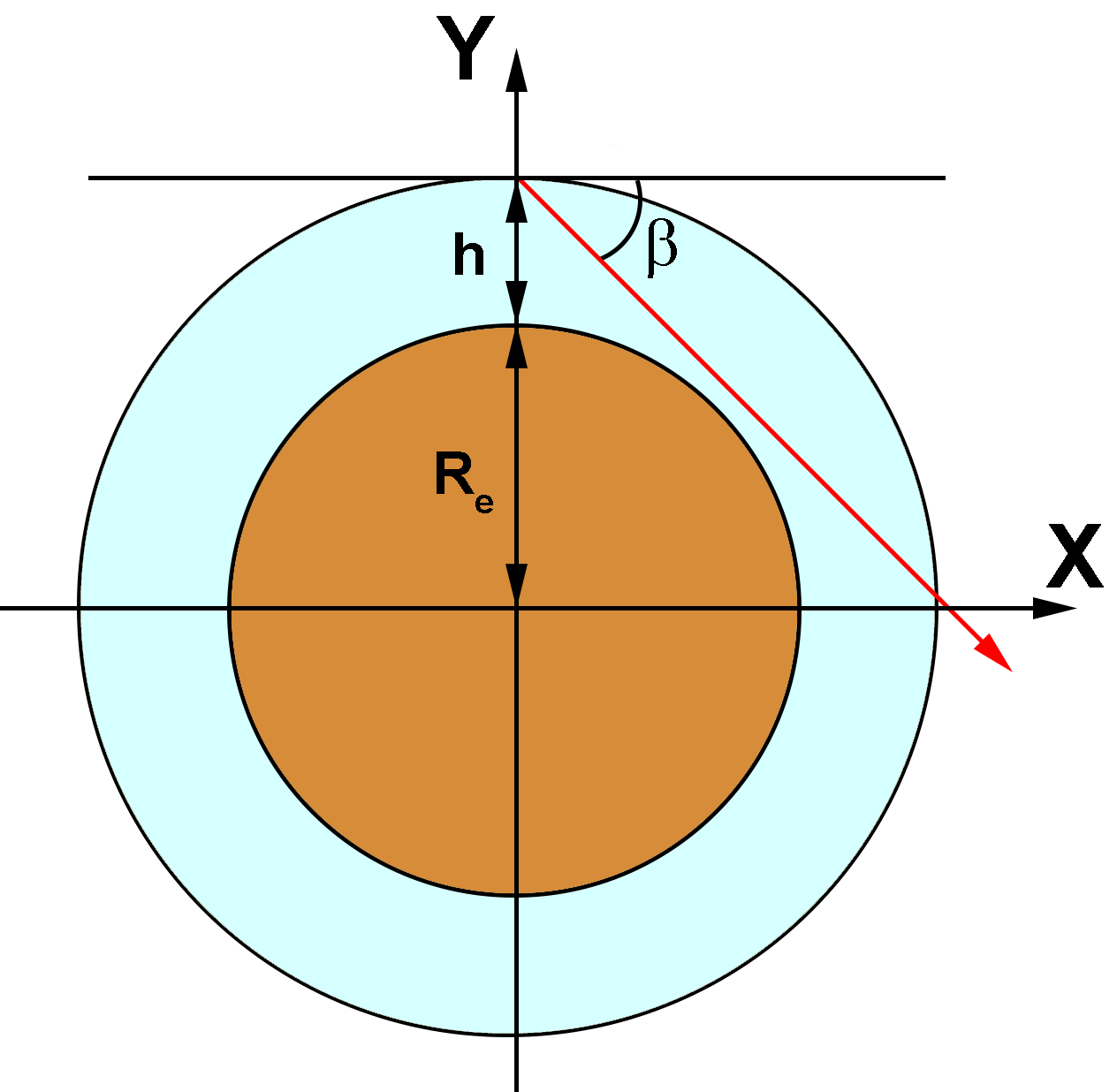}
    \caption{Schematic diagram of the motion of a space body in the Earth's atmosphere  and the angle of entry into the atmosphere ($\beta$) at a given point relative to the $XY$ coordinate system. $R_e$ is the radius of the Earth. Thickness of the atmosphere is exaggerated. The trajectory of SB and its length within the atmosphere are indicated by the line with the arrow.}
    \label{fig1}
\end{figure}

First of all, let us imagine a model explaining the entry of SB into the Earth's atmosphere with respect to the chosen $XY$ coordinate system coincided with  the centre of the Earth and corotating with the rotation of the Earth (Fig.~\ref{fig1}). The altitude of the entry of the SB into the atmosphere is measured from the starting value $h=160$~km, at which the temperature of the SB begins to increase~\citep{Bronshten1981}. This layer of the atmosphere is indicated in Fig.~\ref{fig1} by the value of $h$ (the same parameter denotes the current altitude of SB over the Earth's surface). The angle of entry into the atmosphere relative to the local horizontal line at the altitude $h$ is one of the most important parameters of the problem and is denoted by $\beta$. We denote by $L$ the length along the curved trajectory of SB in the atmosphere, $dL =\sqrt{dX^2+dY^2}$, and $L=0$ corresponds to the entry point of the SB into the atmosphere at $h=160$~km.

\subsection{The equation of motion with a variable mass}

Ballistics of the SB is described by a system of equations, including the equation of motion under the action of applied forces: the force of the aerodynamic drag $F_f$ and the gravitational force ${\bf F}_g=M{\bf g}$~\citep{Stulov1995}:

\begin{equation}
\begin{split}
   M \frac{d{\bf V}}{d t} =- \frac{1}{2} c_d \rho_h V^2 S \frac{\bf V}{V}+M {\bf g}, \\
   \quad{\bf g} =- \frac{GM_e}{r^3} {\bf r}, \   \frac{d {\bf r}}{dt} = {\bf V}
    \end{split}
    \label{eq1}
\end{equation}
Here $M$ is mass, $\bf V$ is velocity of the body relative to the Earth, $t$ is time, $G$ is the gravitational constant, $g$ is the acceleration of gravity, $M_e$ is the mass of the Earth, $c_d$ is the drag coefficient, $\rho_h$ is the density of the atmosphere at altitude $h$~\citep{standart}. $S$ is the area of the body's middle cross-section, $\bf r$ is the current radius vector, and $r$ is the absolute value of the current radius vector (the distance from the SB to the centre of the Earth). The time dependence of $M$ is implied in  Eqs.~\eqref{eq1}, \eqref{eq3} and \eqref{eq4}. 

We note that the contribution of the lifting force to the ballistic of the SB is also neglected in Eq.~\eqref{eq1} because we assume that the shape of the SB is close to spherical. 
The Coriolis and centrifugal forces in the rotating reference frame are negligible for fast moving SBs compared to the aerodynamical forces from stratospheric winds, which we also neglect here because SB moves much faster than the wind speed.

In accordance with the existing ideas  e.g.~\citet{Stulov1995}, the main contribution to the force of aerodynamic drag is made by the difference in pressures between the frontal and rear parts of the SB surface (low pressure cavity forms near the rear surface). The expression for the  force of  aerodynamic drag corresponds to the first term in Eq.~\eqref{eq1}. 
The area $S$ depends on the current mass and size of SB (Eq.~\eqref{eq3} below). In Eq.~\eqref{eq3} the initial values of the cross-sectional area and the mass of the SB are indicated as $S_0$ and $M_0$
\begin{equation}
S(t,M)=S_0 \left( \frac{M(t)}{M_0}\right)^{\mu}.
\label{eq3}
\end{equation}

We apply  the simplified approach of isotropic  loss of material from the SB surface, which corresponds to the case $\mu={2}/{3}$. Determining the exponent $\mu$ is the subject of a separate study \citep{Bronshten1981}  taking into account the complexity of the problem in  general case.
The dependence $S(t,M)$ is used in the numerical solution of the differential equation\eqref{eq1} at each time iteration step  together with the change of the SB mass (see Eq.~\eqref{eq4} below).

The  value of the SB drag  coefficient~\citep{Kutateladze1990} in  general case depends on the Reynolds number $Re=\frac{VR}{\nu}$, where $R$ is the radius of SB, $\nu$ is the kinematic viscosity of the air. At the velocity in the range $10<V<40$~km/s, radius of SB is of the order of several dozens of metres, kinematic viscosity of air at a trajectory altitude is of the order of $5\cdot10^{-5}$~m$^2$/s and greater. Therefore, the Reynolds number exceeds $10^{10}$~\citep{Bronshten1981}. According to~\citep{Kutateladze1990,Spearman1993,Zhdan2007}, if the Reynolds number exceeds the value $1.5\cdot10^5$, then a sharp drop of $c_d$ to 0.1 with subsequent raise on the average up to  0.9--1 takes place. In our case the value of $c_d$=0.9 corresponds to the most extreme conditions for the motion of spherical SB in the atmosphere. 

\subsection{The loss of the space body mass and the heat transfer with the shock wave boundary layer}
Equation~\eqref{eq4} describes the loss of SB mass when moving in the atmosphere
\begin{equation}
\frac{dM}{dt}=-\frac{c_h \rho_h V^3 S}{2H}.
\label{eq4}
\end{equation}
Here $H$ is the specific heat of sublimation of the SB material, $c_{h}$ is the coefficient of radiation heat transfer, defined as the fraction of the kinetic energy of the oncoming stream of molecules which goes into sublimation of the SB material~\citep{Bronshten1981}. 

Mass loss of SB occurs due to heating to the temperature much higher than the melting point~\citep{Stulov1995}. In our case the main contributor to this heating is radiant heat transfer between the SB and the boundary layer of the shock wave, whose temperature reaches several thousand degrees close to the surface of the SB~\citep{Bronshten1981}. One of the most difficult problems in calculating the radiant heat transfer is the determination of the radiant heat transfer coefficient ($c_h$). Its magnitude is affected by the velocity of motion in the atmosphere, flight altitude, air density, temperature of the boundary layer and the nature of the processes in the boundary layer (dissociation and ionization of air molecules), the degree of blackness of the radiating and absorbing surface, etc. According to the available data, depending on the altitude and the SB velocity, the values of the radiant heat transfer coefficient lie in the range $0.01\leq c_h\leq 0.1$~\citep{Johnson2018,Andruschenko2013,Svetsov1995} (the maximum value is reached at the altitude of about 10~km). Taking into account  that the mass loss rate reaches a maximum at $c_h=0.1$, this value will be used in further calculations. According to \citep{Svetsov1995}, this value gives the best fit with observational data.  The degree of blackness for the radiating and absorbing surface was taken equal to 1 in the calculations because of the high temperature of the surfaces and the formation of a dense high-temperature plasma~\citep{Bronshten1981}. The solution of the system of differential equations~\eqref{eq1} and \eqref{eq4} together with algebraic equation \eqref{eq3} was carried out  by the explicit Runge-Kutta method of the fourth order.

Note that the employed model does not involve the process of SB fragmentation~\citep{Barri2010,Stulov2001,Dudorov2015}, since the initial dimensions of the SB are taken to be quite significant (from 50 to 200 metres) as well as moderate velocities, when most of the SB remains intact, despite extreme external influences. First of all, maximum resistance to fragmentation is characteristic of iron SBs that is associated with the high homogeneity of their internal structure. In contrast to the iron SBs, the internal structure of stone and ice SBs is heterogeneous with an abundance of numerous microcracks. The results of study of the conditions for the fragmentation of iron SBs will be presented in our next paper.

The mass loss we denote by the term "ablation", which includes two processes -- the low-temperature blowing off a liquid film from the SB surface (at the temperature about 1000~$^\circ$C) with the formation of small  droplets. These droplets are typical for slow fall of small SBs or their fragments at the final stage of the flight in atmosphere. The second process is the high-temperature sublimation of material occurring when the surface temperature exceeds several thousand degrees. In this case a mass loss occurs in the form of vapors of single atoms and their ions. In  the conditions under consideration the employed model includes the sublimation as dominant process responsible for the mass loss at high velocities -- over 11.2 km/s.  

\section{Results and discussion}
\begin{figure}
\centering
	\includegraphics[width=6cm]{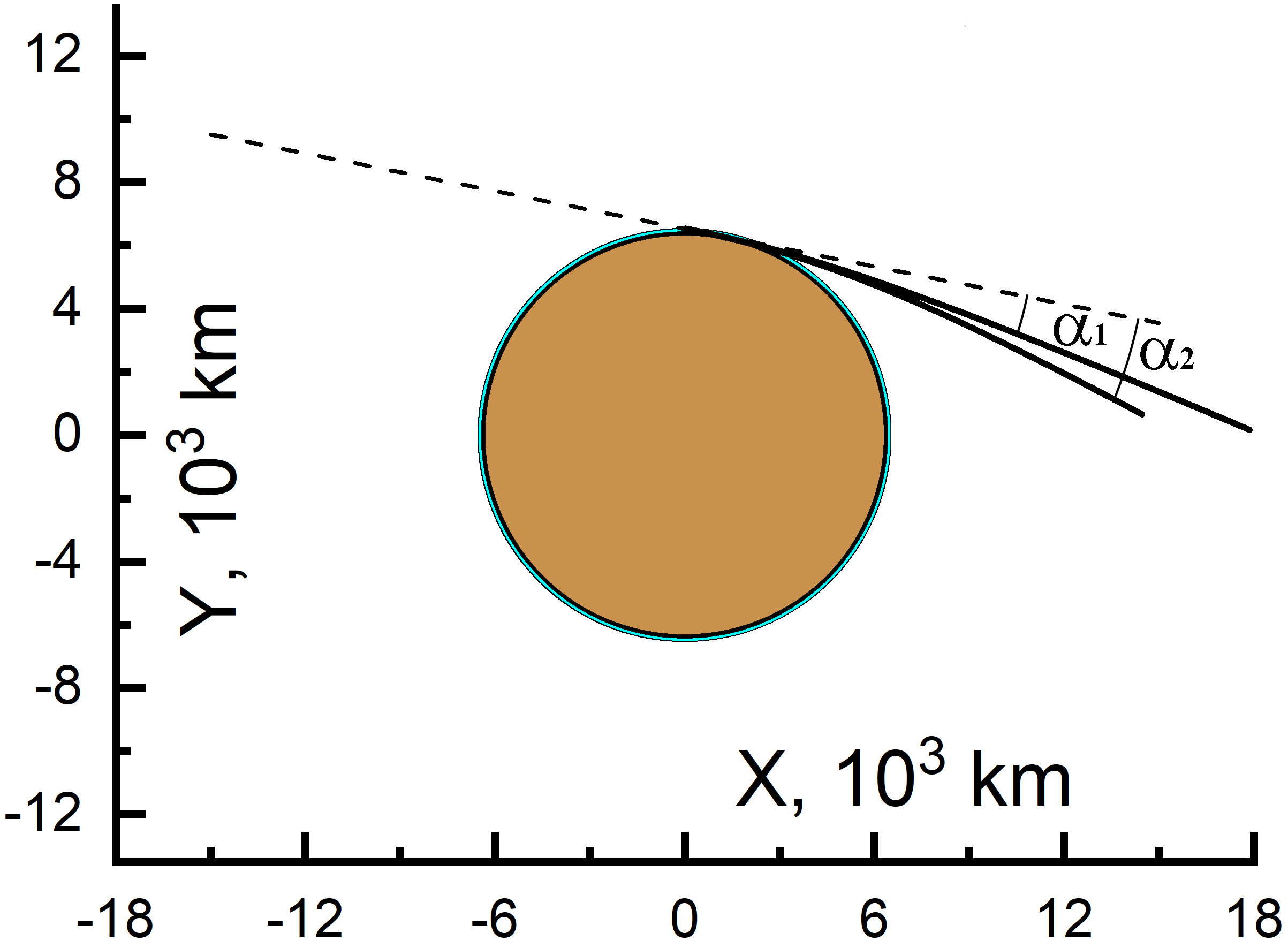}
    \caption{Changes in the trajectory of SB during a through passage via the atmosphere. The SB parameters are radius $R=50$~m, the velocity of entry into the atmosphere is 20~km/s, the minimum altitude is 11 km. 
The angle of deflection $\alpha_1=11.25^\circ$ at $c_d$=0 and $\alpha_2=16.9^\circ$ at $c_d$=0.9. The trajectory lengths corresponds to the time moment 1000 s after  the entrance into the atmosphere.}
    \label{fig2}
\end{figure}

As a typical example of our calculations, Fig.~\ref{fig2} illustrates the results of calculating the trajectory of the spherical iron SB with a radius $R=50$~m entering into the atmosphere at 20~km$/$s when passing through it at the entry angle $\beta=11.2^\circ$ and  the minimum altitude of 11~km. As can be seen from this figure, the perturbation of the trajectory of the SB deviates it from the initial direction by the angle $\alpha_1=11.25^\circ$ when neglecting the aerodynamic drag effect and $\alpha_1=16.9^\circ$ when we take it into account. These results demonstrate the  significant effect of aerodynamic drag on the SB trajectory. 

\begin{figure}
\centering
	\includegraphics[width=5cm]{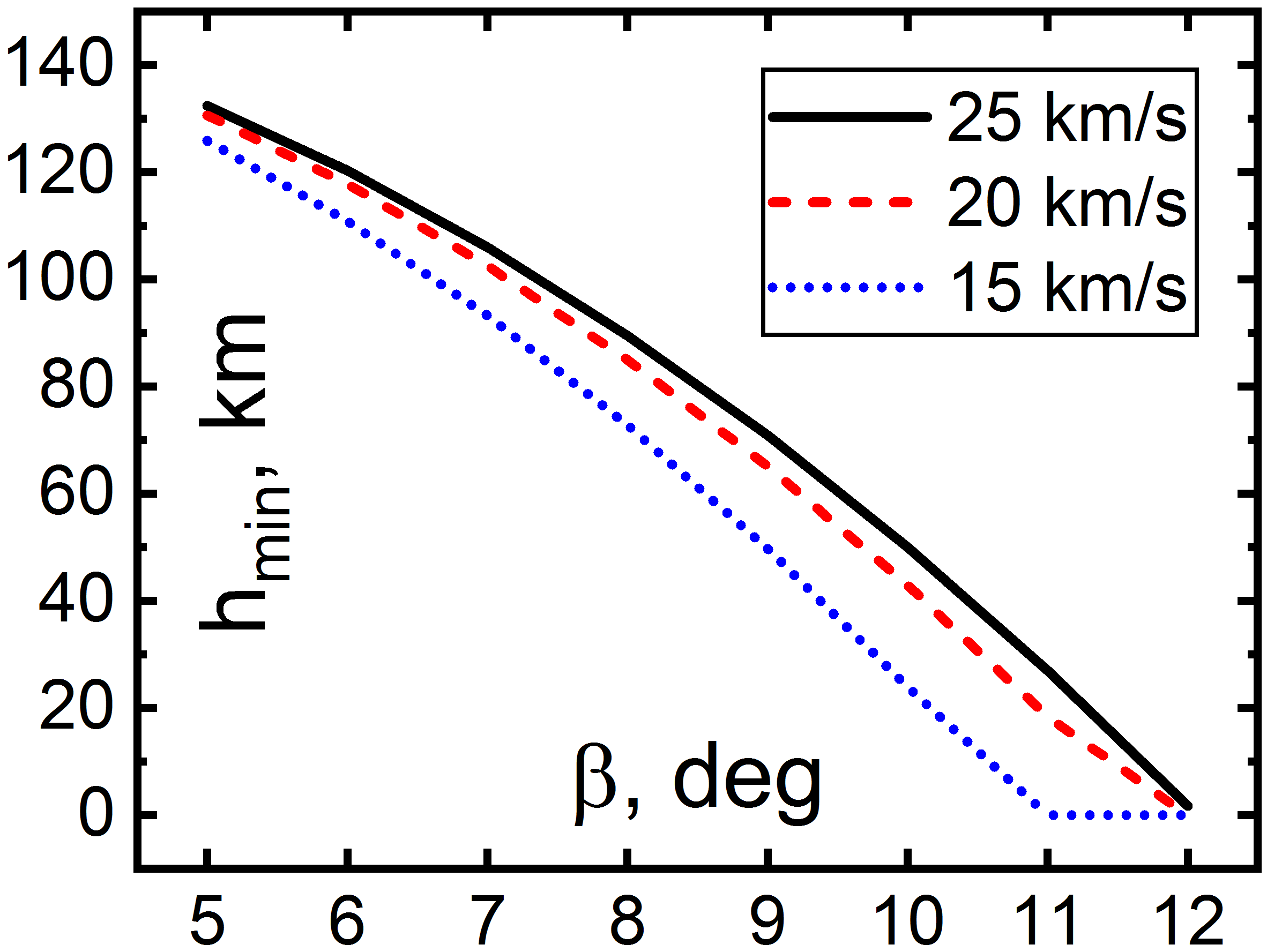}
    \caption{Dependence of the minimum trajectory altitude on the entry angle for the R=100 m iron SB at different velocities.}
    \label{fig3}
\end{figure}

Fig.~\ref{fig3} shows the dependence of the minimum trajectory altitude for iron  SB with R=100 m on the  angle of entry into atmosphere for three values of velocities. Calculations did not reveal significant differences for the SBs with radii  R=100, 50 and 25~m.

\begin{figure}
\centering
	\includegraphics[width=5cm]{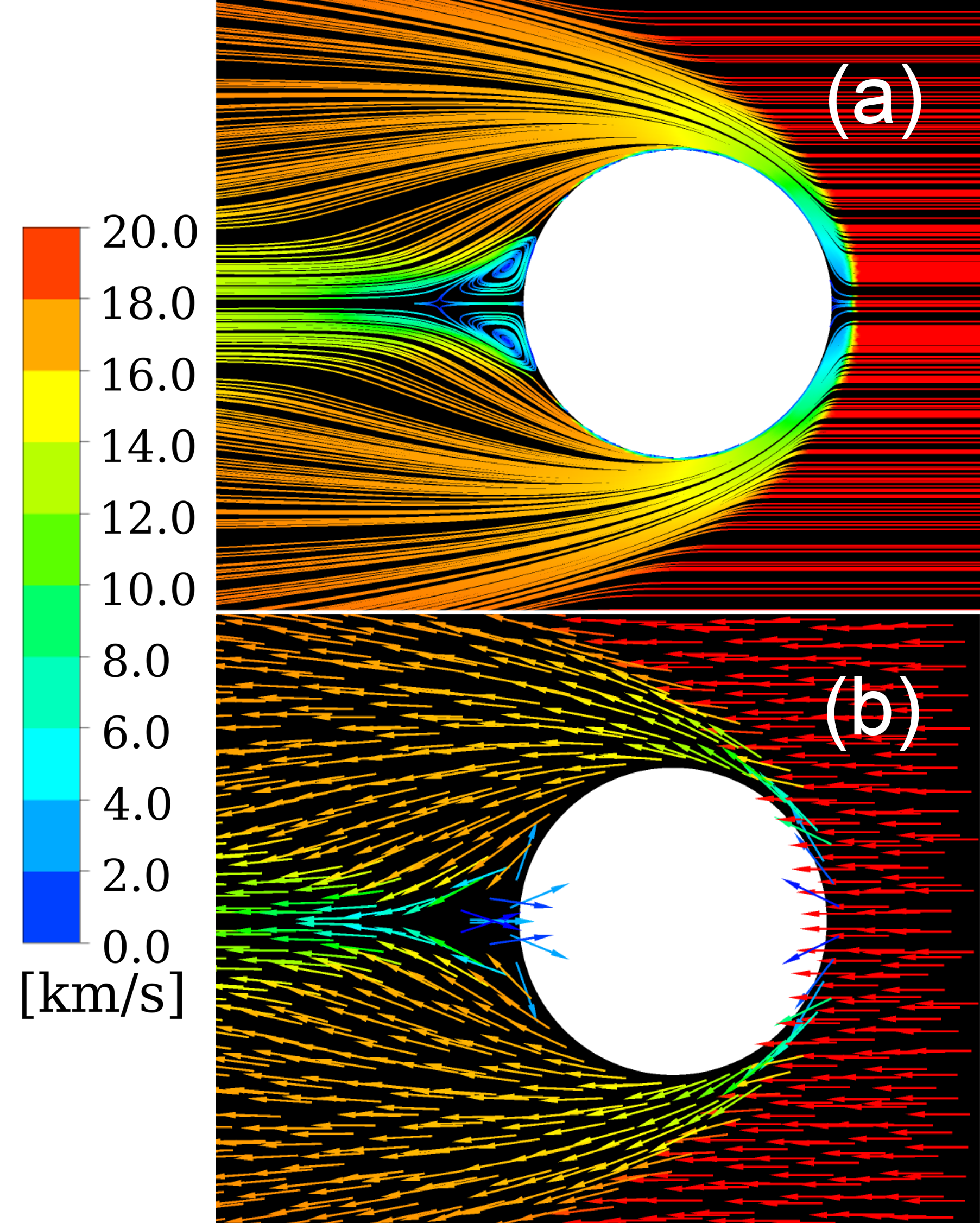}
    \caption{(a)~--- distribution of the velocity modulus around spherical SB with radius $R=100$~m; (b)~--- distribution of the velocity vectors.}
    \label{fig4}
\end{figure}

The complex pattern of  aerodynamic fluxes around a spherical SB, vortex formation and stagnation zone in its rear part are shown in  Fig.~\ref{fig4}. This figure  demonstrates the extreme conditions for the passage of a spherical SB through the dense layers of the atmosphere  as well as  clearly shows the conditions for the occurrence of the aerodynamic drag taking into account the structure of fluxes in the stagnation zone. 
The results illustrated in Fig.~\ref{fig4} 
were obtained with the software package ANSYS Fluent~\citep{ansys}. This package is a universal software system of the finite element method applied for solving various problems in aero- and hydrodynamics~\citep{ansys}. To calculate the pressure distribution at the surface of an SB with the effect of air compression due to pressure-density dependence the Fluent ``ideal gas'' model was used with the laminar flow regime. Zero static pressure was set to absolute pressure 26500~Pa at the altitude of 10 km~\citep{standart}. The default air temperature was set to 219~K.

\begin{figure}
\centering
	\includegraphics[width=7cm]{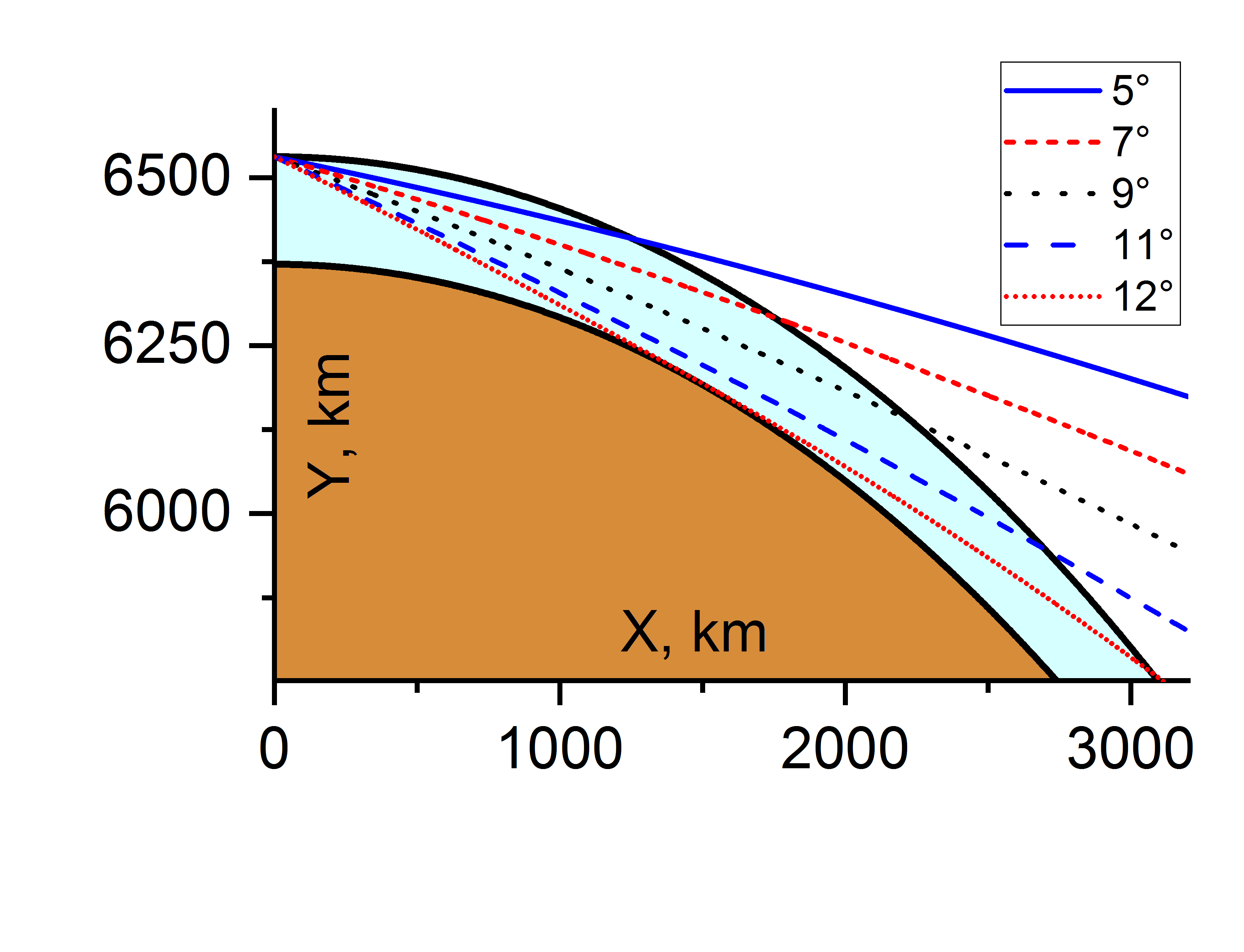}
    \caption{The trajectories of the passage of SB in the atmosphere at the entry angles ($\beta$) from $5^\circ$ to $12^\circ$. Material is iron, radius of SB is $R=100$~m, the initial velocity is 25~km/s. }
    \label{fig6}
\end{figure}

Fig.~\ref{fig6} shows the trajectories of an SB at different angles of entry into the atmosphere  corresponding to the  passage of SBs through the atmosphere at different minimum altitudes over the Earth's surface. 

In this section we present the results of calculations for bodies of several sizes, consisting of iron, stone and ice. Calculations were carried out for the values of the radii of the  SBs with $R=25$, 50 and 100~m and for three materials: for iron, the specific heat of sublimation of iron is $H=6380$~kJ/kg~\citep{Luchinsky1985}, for stone with $H=3965$~kJ/kg for a specimen of lunar rock~\citep{Ahrens1971} and with $H=9300$~kJ/kg~--- for quartz~\citep{Chirikhin2011}, and for water ice with $H=2853$~kJ/kg~\citep{Voitkovskiy1999}. In calculations with a stone SB, we used the larger value of $H$ for quartz, since SiO$_2$ is the basis of many natural minerals. The use of the smaller value of $H$ (3965~kJ/kg for lunar rock) results in increase of the  mass loss of stone SBs.

\begin{figure}
\centering
\begin{tabular}{cc}
	\includegraphics[height=3.0cm]{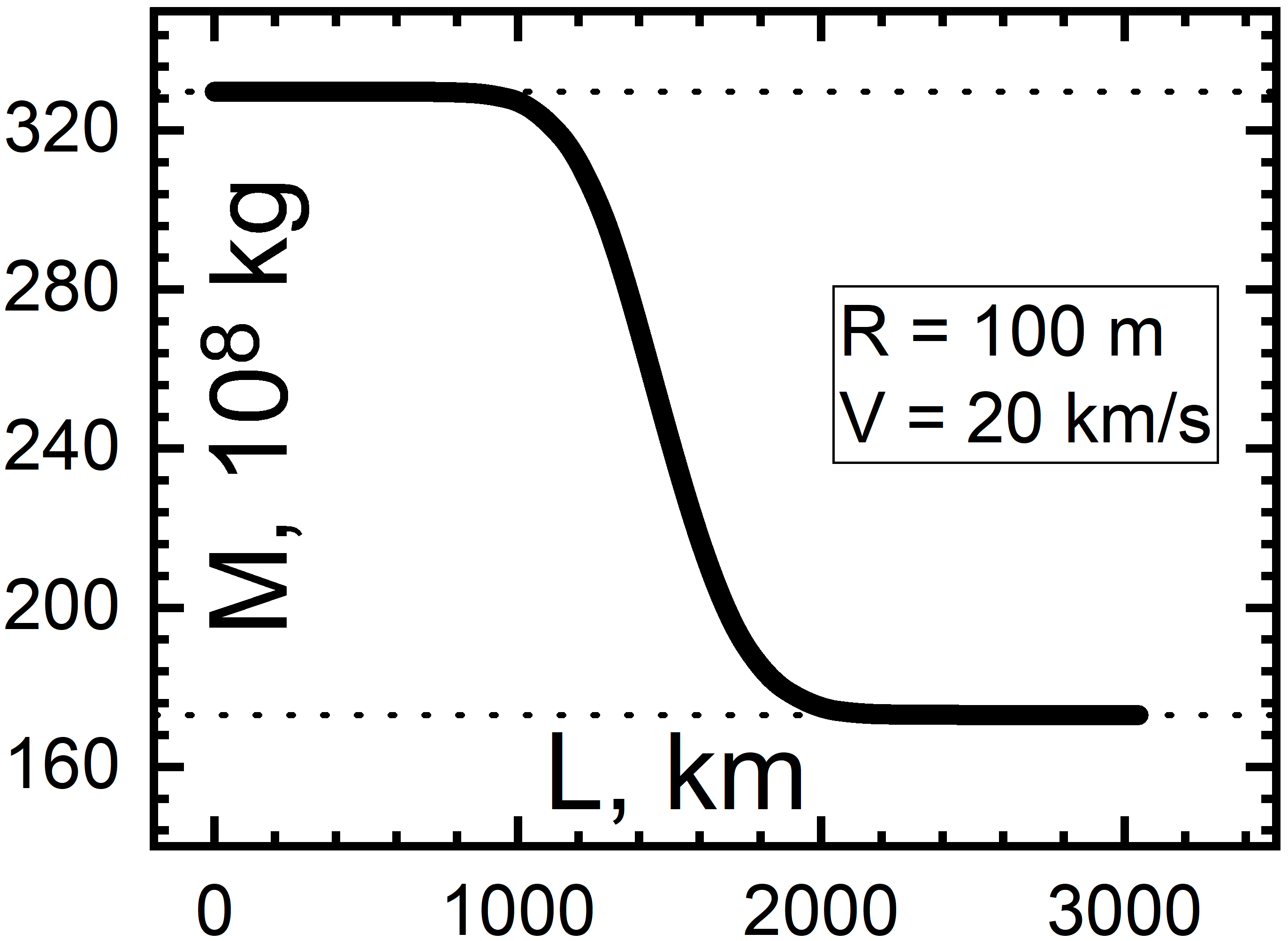}&
	\includegraphics[height=3.0cm]{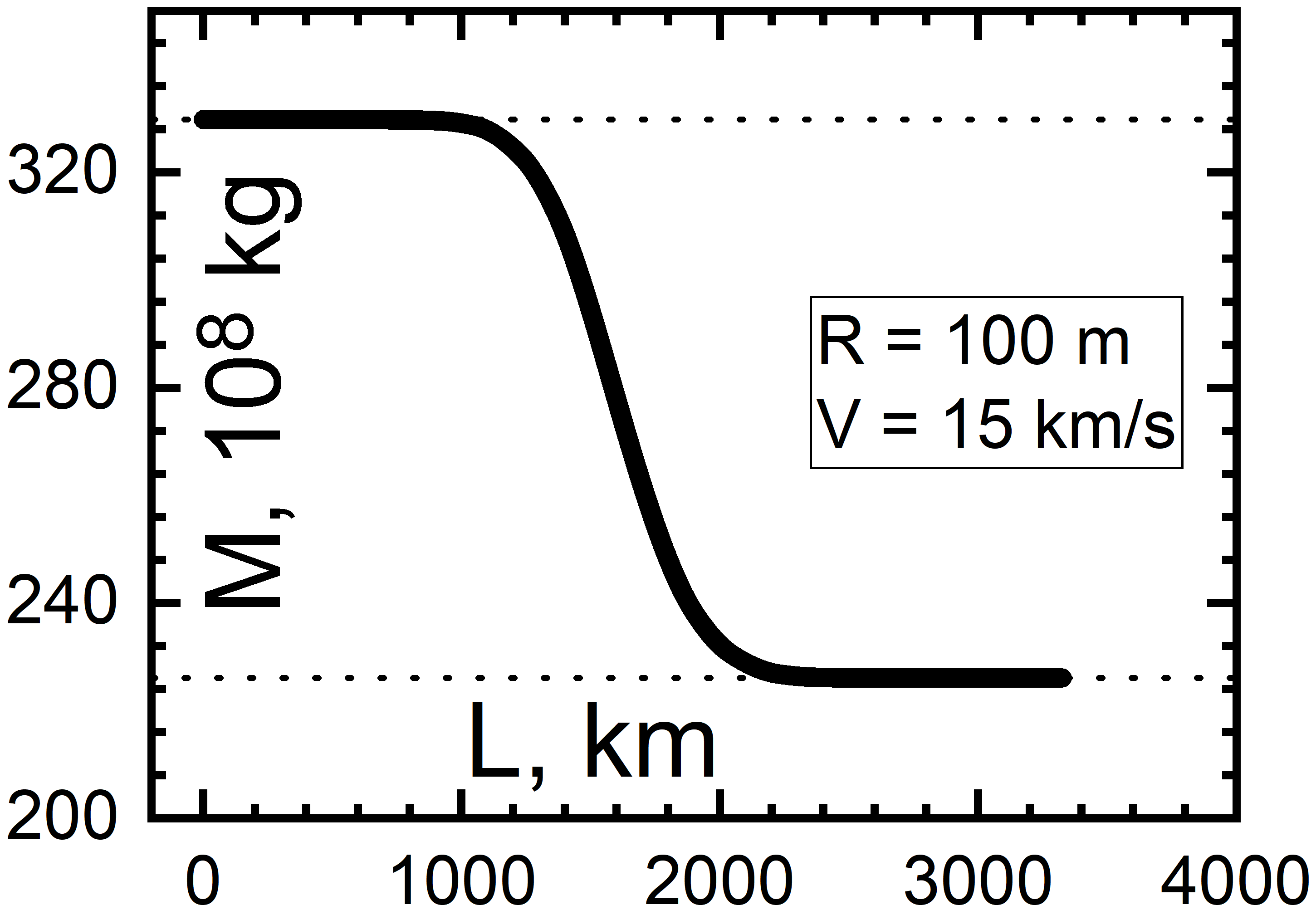}\\
(a)&(b)\\
	\includegraphics[height=3.0cm]{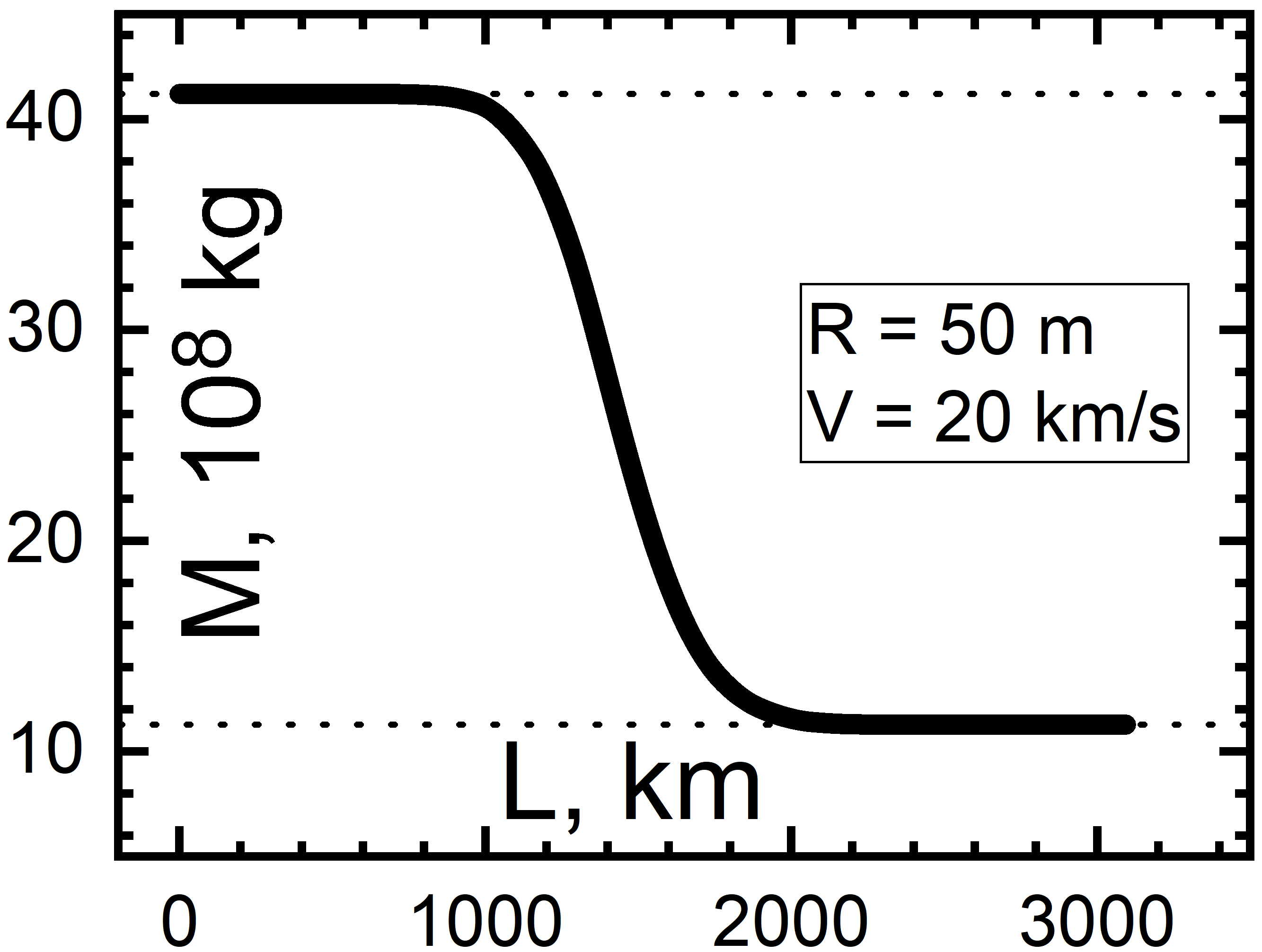}&
	\includegraphics[height=3.0cm]{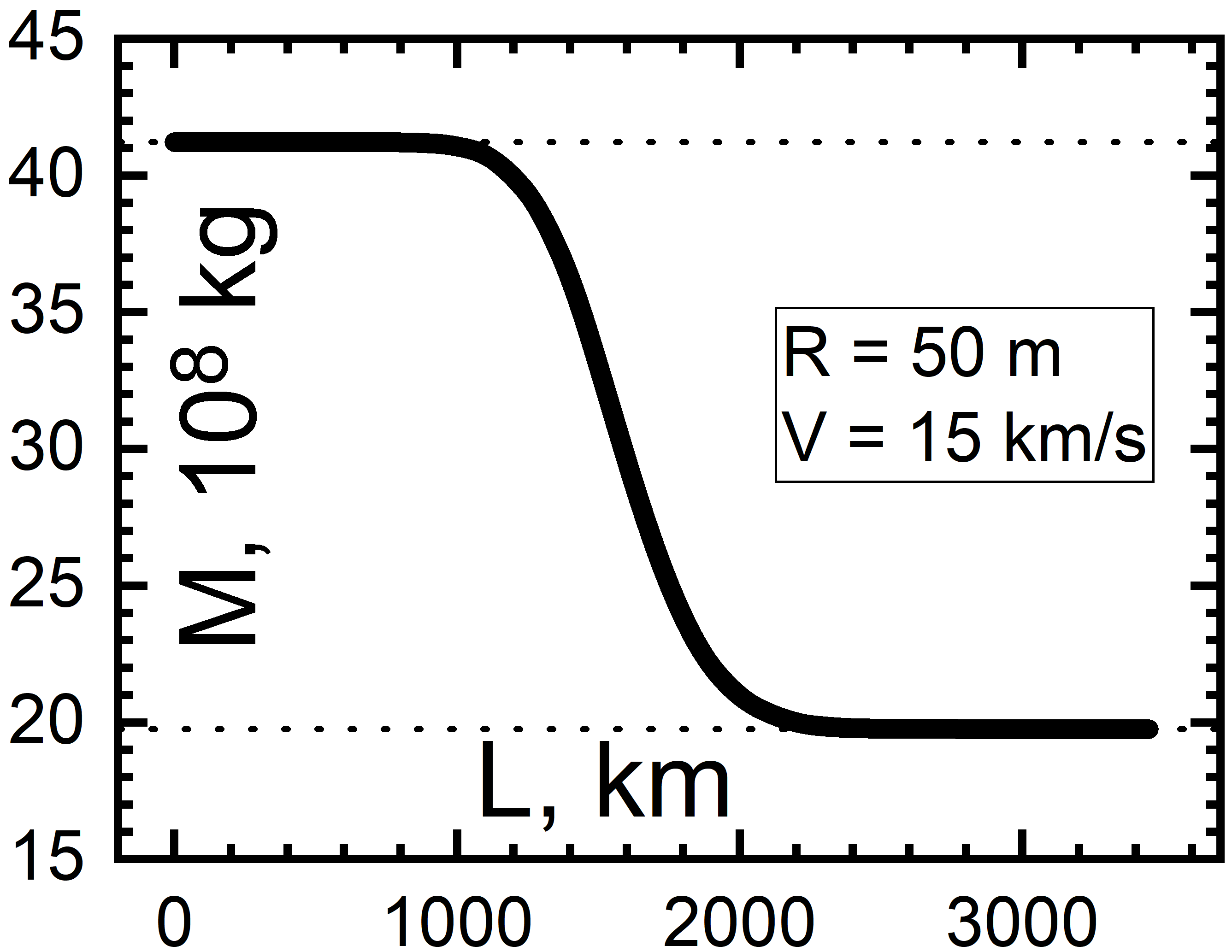}\\
	(c)&(d)\\
	\includegraphics[height=3.0cm]{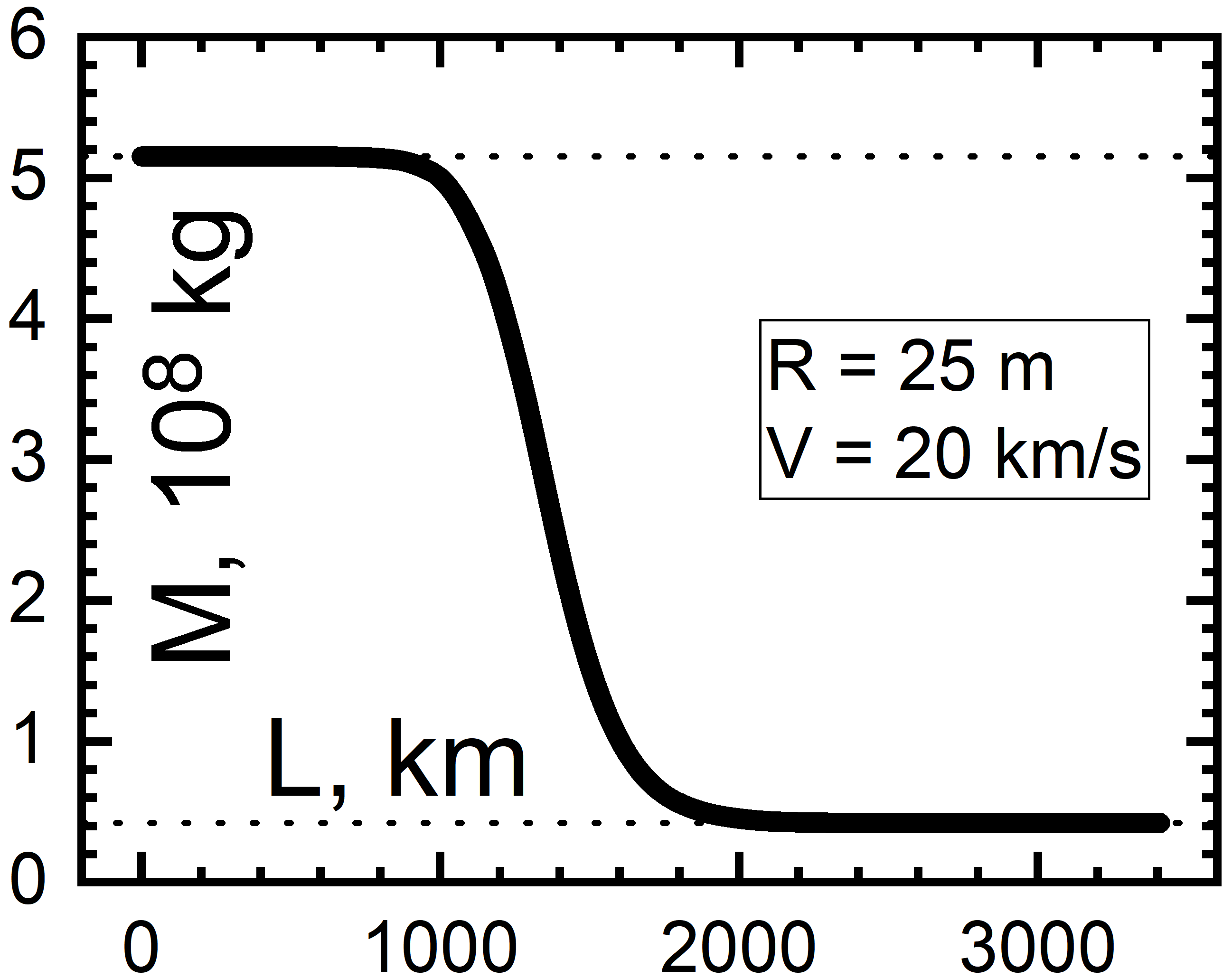}&
	\includegraphics[height=3.0cm]{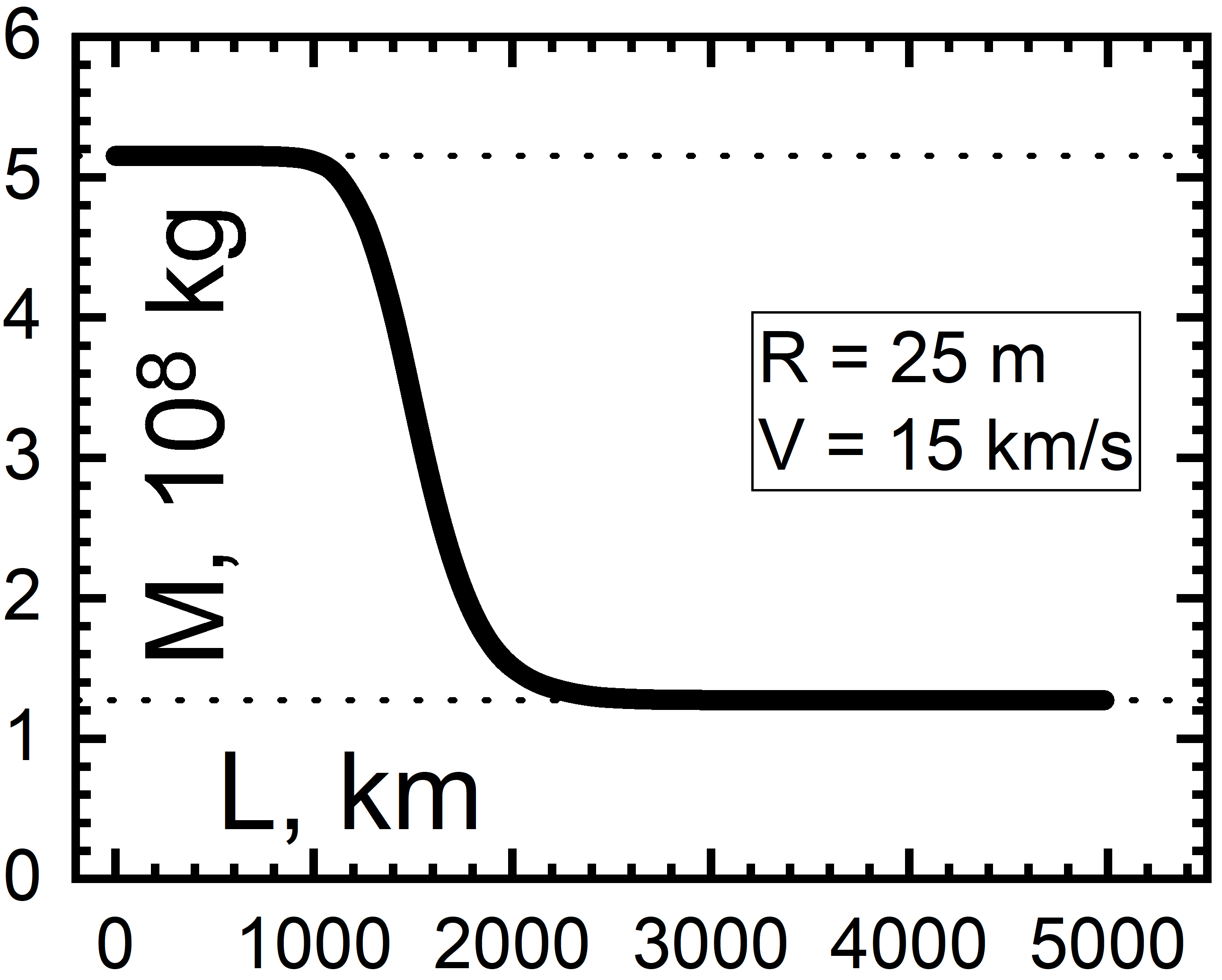}\\
		(e)&(f)
\end{tabular}
\caption{Change in the residual mass of the iron SBs ($M(L)$) along the trajectory through the atmosphere 
at two values of initial velocities and for three SB radii. $L$ is the 
length along the trajectory measured from the entry point into the atmosphere at the altitude $h=160$~km. The curves end when the SB reaches the exit point from the atmosphere at the same altitude $h=160$~km. The minimum altitude $h_{min}$=11~km, $c_d=0.9$. 
}
\label{fig7}
\end{figure}

\begin{figure}
\centering
\begin{tabular}{cc}
	\includegraphics[height=3.0cm]{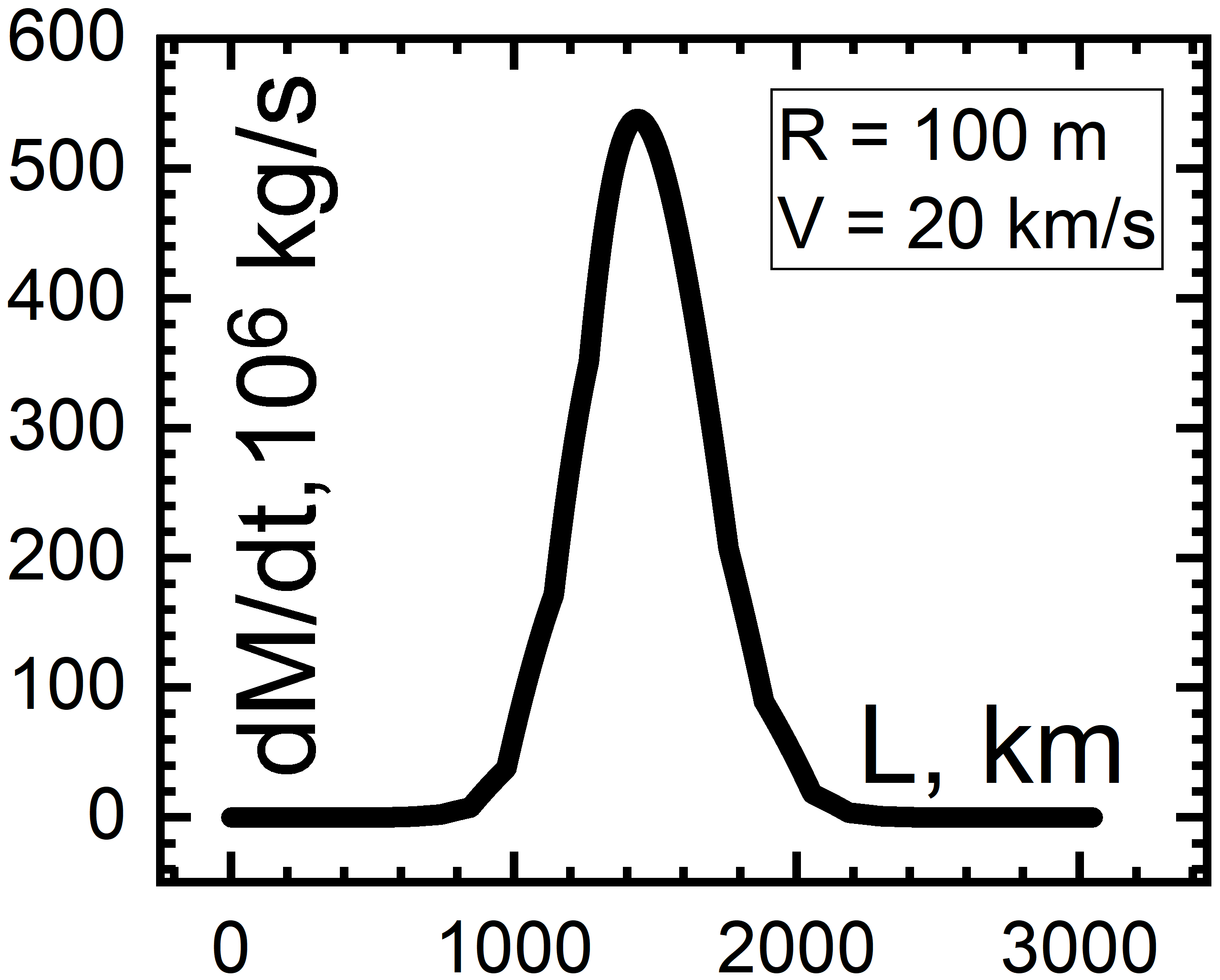}&
	\includegraphics[height=3.0cm]{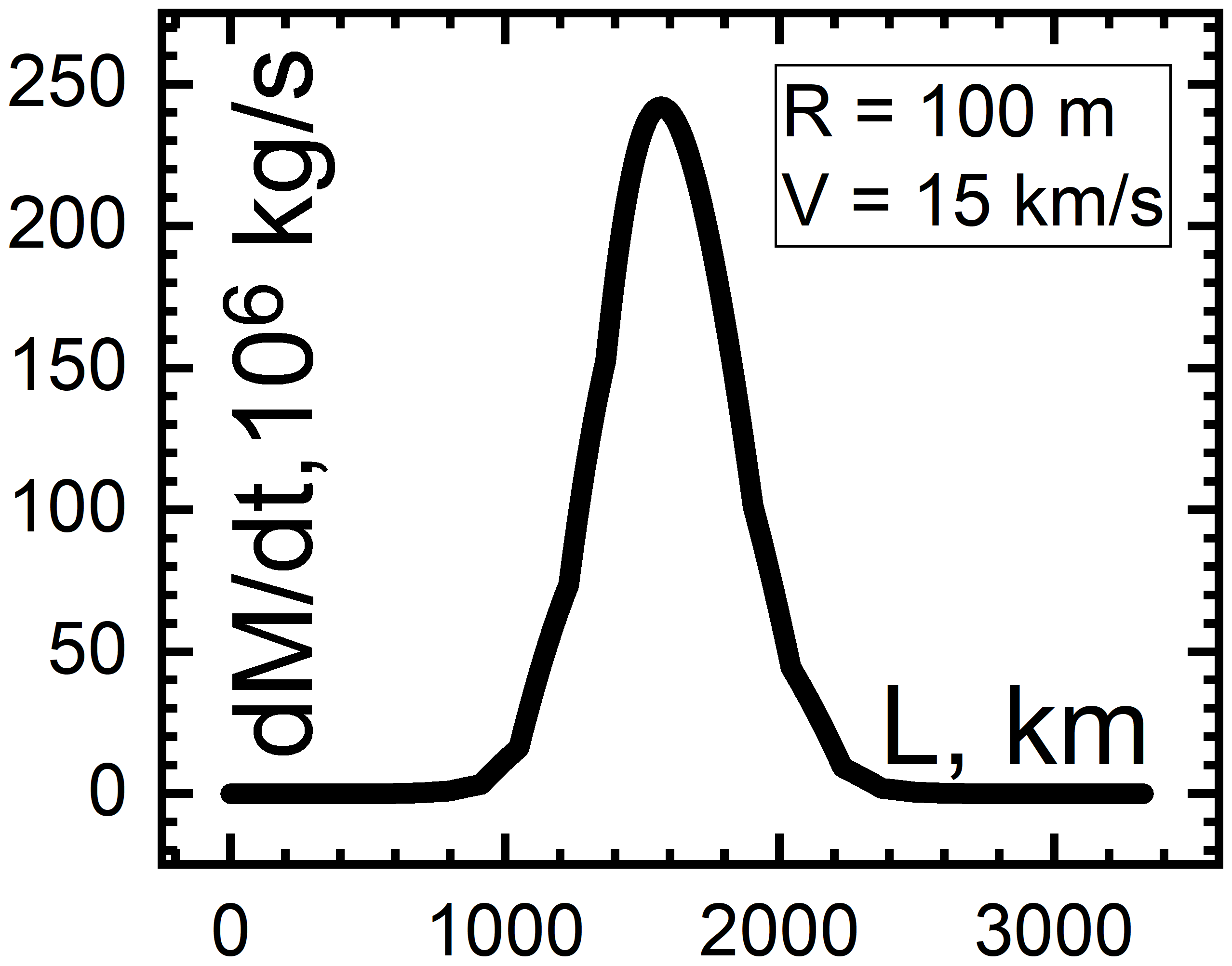}\\
	(a)&(b)\\
	\includegraphics[height=3.0cm]{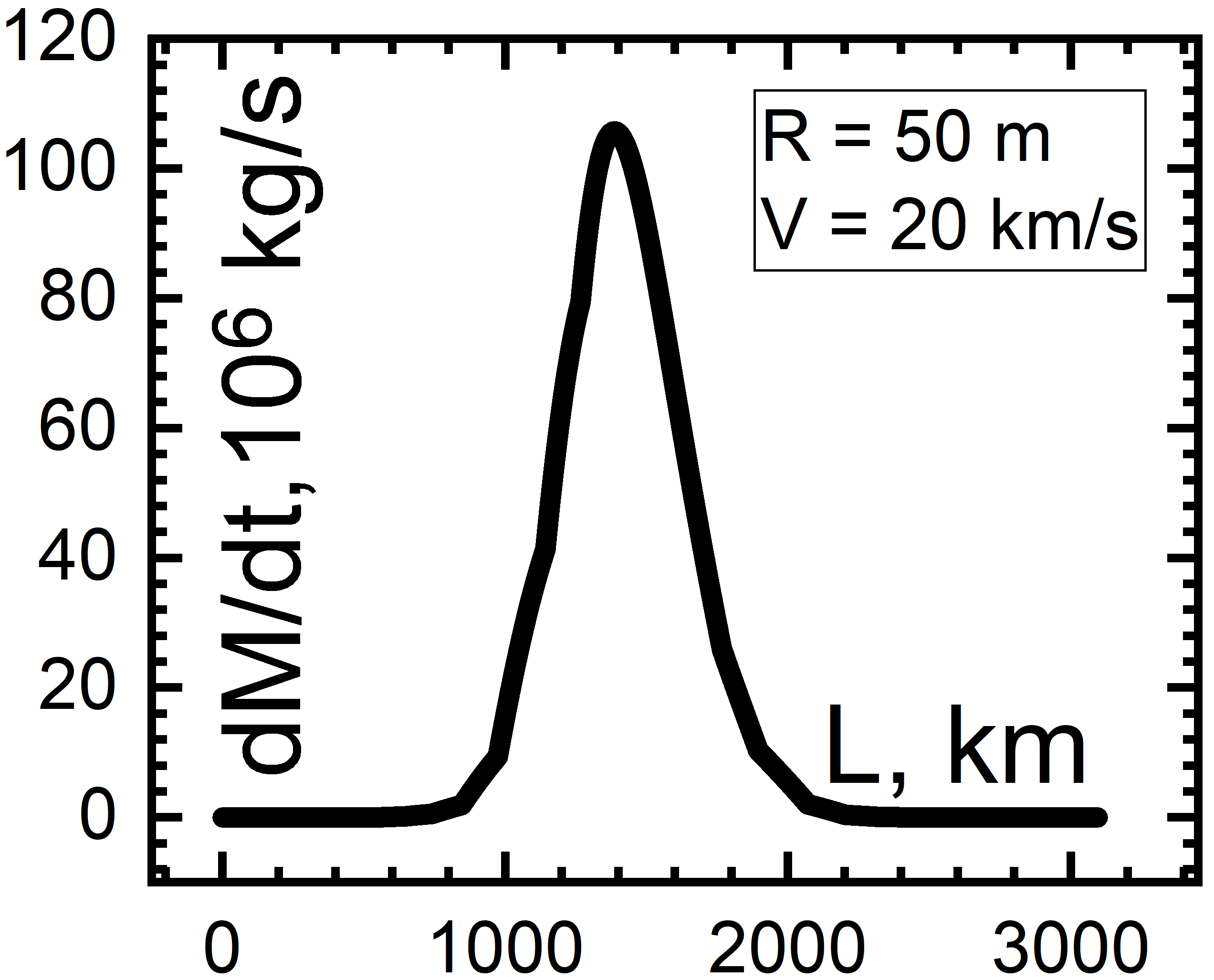}&
	\includegraphics[height=3.0cm]{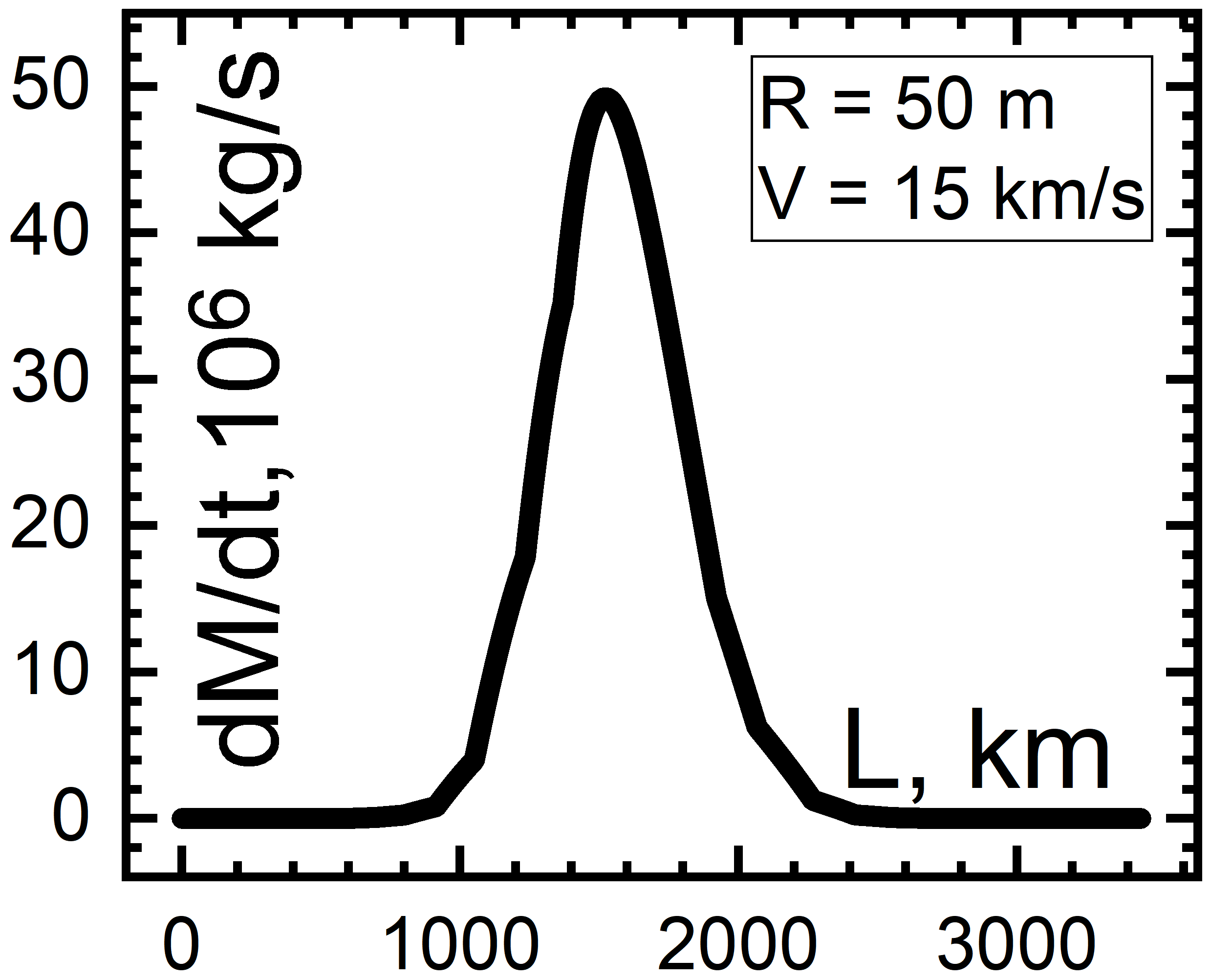}\\
	(c)&(d)\\
	\includegraphics[height=3.0cm]{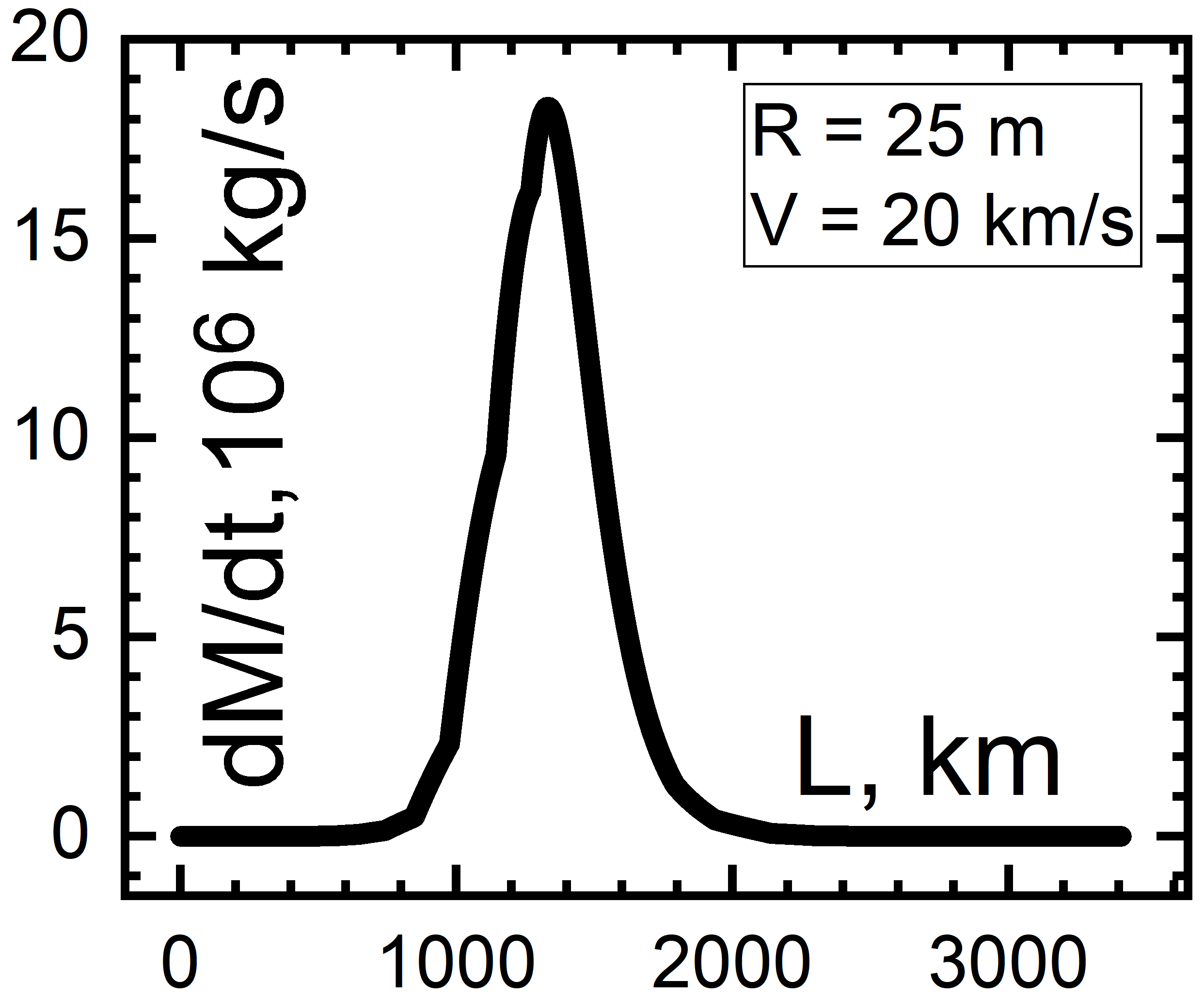}&
	\includegraphics[height=3.0cm]{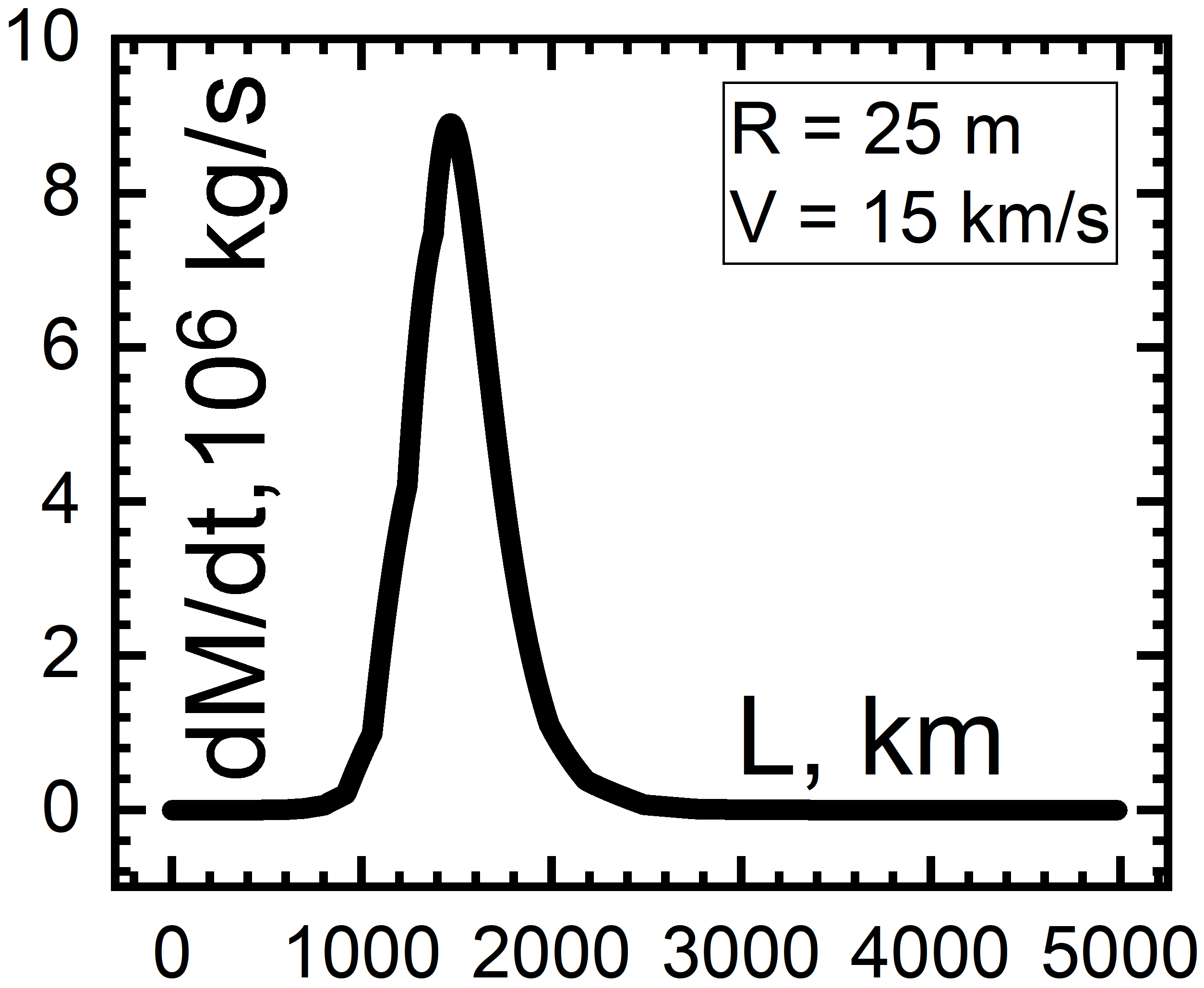}\\
	(e)&(f)
\end{tabular}
\caption{Change in the rate of the mass loss of the iron SBs  ($dM(L)/dt$) along the trajectory through the atmosphere at two values of initial velocities and for three SB radii. The curves end when the SB exits from the atmosphere at the altitude $h=160$~km. The minimum altitude $h_{min}$=11~km, $c_d=0.9$.  
}
\label{fig8a}
\end{figure}

Fig.~\ref{fig7} and~\ref{fig8a}   illustrate  the mass loss of an iron SB and its  rate  when it is moving through the atmosphere at different velocities and at minimum trajectory altitude of 11 km. This  altitude lies in the range of generally accepted values of the minimum altitude for the Tunguska space body: 10 - 15 km \citep[e.g.,][]{Bronshten1981}. These dependencies are affected both by the velocity of SB, which increases the loss of material, and the time of flight through the atmosphere, which reduces this loss.

Besides less loss of mass at lower velocity (15 km/s) we can see the effect of lengthening of trajectory in atmosphere from 3000 to 5000 km which takes place at smaller size of SBs as well (Fig.~\ref{fig7}f). 

\begin{figure}
\centering
	\includegraphics[width=5cm]{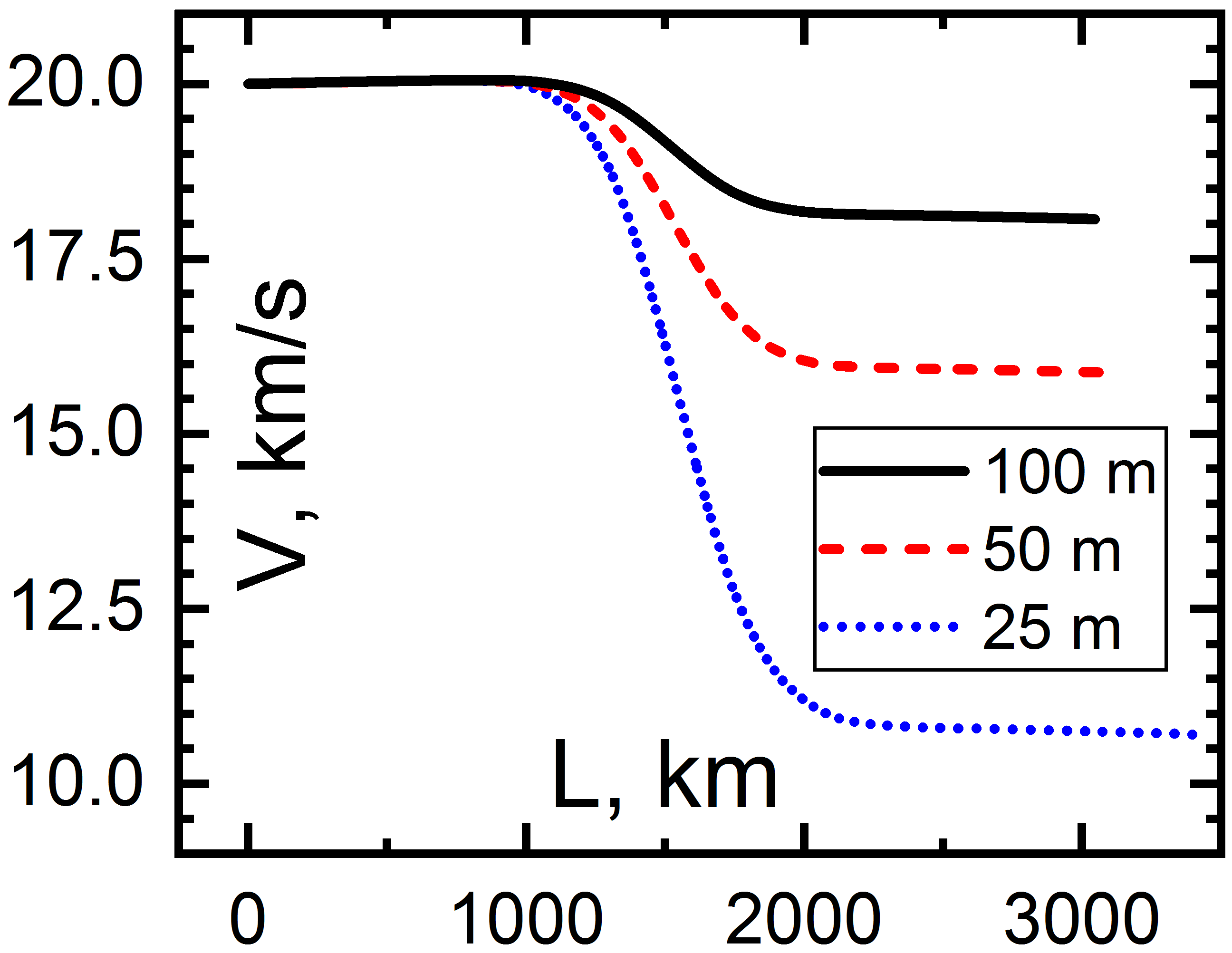}
    \caption{Changes in the velocity of the iron SBs along the trajectory when passing through the atmosphere for the radii $R=100$~m, 50~m, and 25~m at the minimum altitude of 11~km ($c_d=0.9$). The curves end when the SB exits from the atmosphere at the altitude $h=160$~km.}
    \label{fig9a}
\end{figure}

\begin{figure}
\centering
	\includegraphics[width=7.5cm]{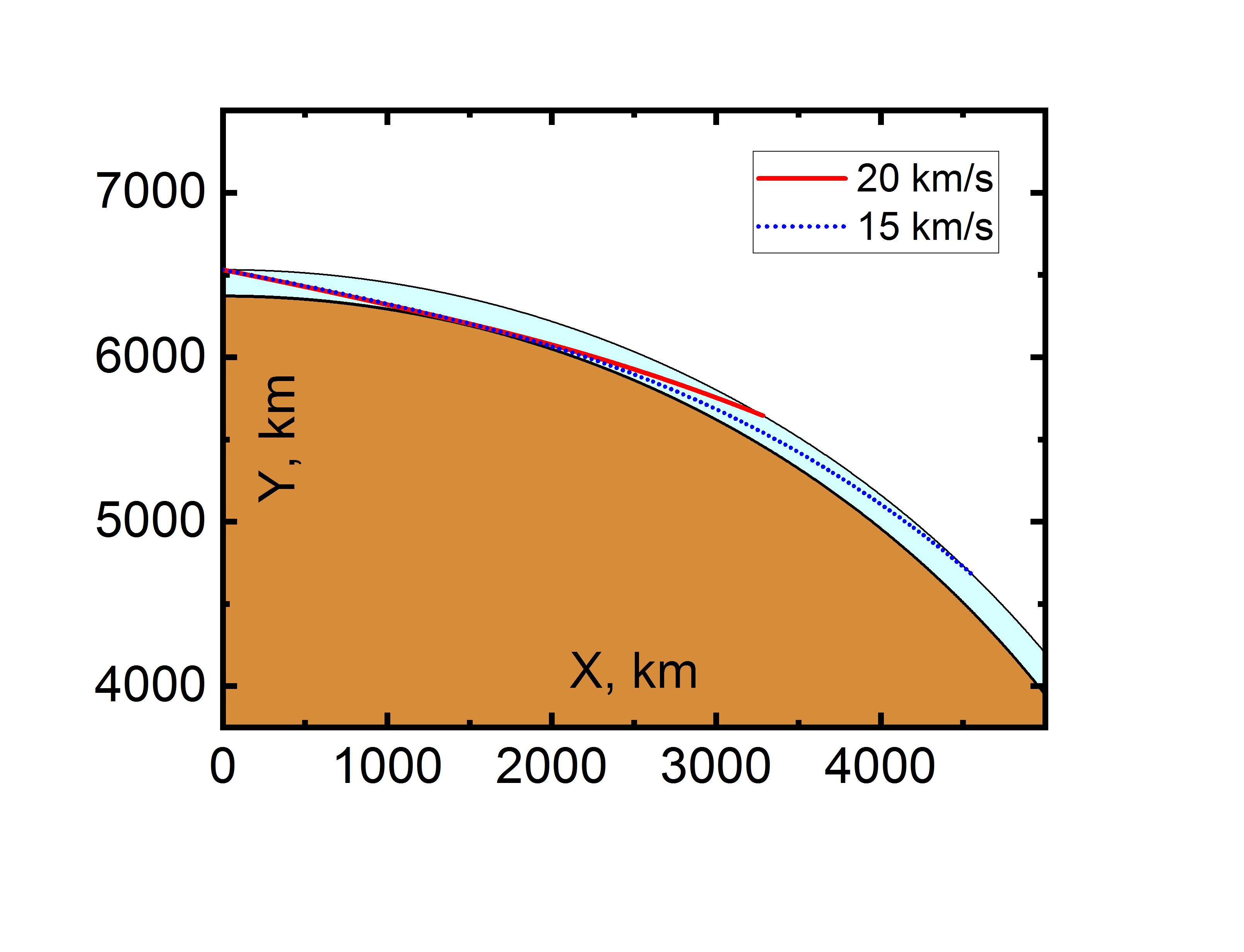}
	\caption{Difference in the iron SB trajectories  with radius $R=25$~m   when passing through the atmosphere with two values of initial velocity and the minimum altitude of 11~km ($c_d=0.9$).} 
    \label{fig10}
\end{figure}

The main results of our calculation are presented in Table~\ref{table1} which can be considered as upper estimates of the residual masses of SBs for different sizes and materials after through passage of the Earth's atmosphere  for the trajectory length of 3000~km.
\begin{table*}
	\centering
	\caption{The ratios  of the preserved mass of SB $M_{out}$ to the initial mass $M_{in}$ for different materials at different initial velocities $V$ and various initial sizes $R$. Calculations are performed for minimum altitude $h_{min}$=11~km, except for the stone SB with $R$=25 m $(h_{min}$=18 km), and for the ice SBs with radii $R$=100, 50 and  25 m with corresponding $h_{min}$= 18, 23 and 28 km.}  
	\label{table1}
	\begin{tabular}{|l|c|c|c|c|c|c|c|c|c|} 
		\hline
		Material&\multicolumn{3}{c}{Iron}&\multicolumn{3}{c}{Stone}&\multicolumn{3}{c}{Ice}\\
		\hline
        $R$, m&100&50&25&100&50&25&100&50&25\\
		\hline
		$M_{out}/M_{in}$ ($V = 15\text{ km/s}$)&0.69&0.48&0.25&0.49&0.32&0.3&0.017&0.014&0.01\\
        \hline
        $M_{out}/M_{in}$ ($V = 20\text{ km/s}$)&0.52&0.27&0.08&0.298&0.11&0.1&0&0&0\\
        \hline
	\end{tabular}
\end{table*}
As can be seen from the Table~\ref{table1}, the maximum  fraction of the preserved mass is observed in the iron SBs with the radius of 100~m at lower initial velocities. When the velocities grow, the residual mass considerably falls. Similar tendencies are observed for the stone SBs (without regard to their fragmentation) with a significantly greater relative  mass loss. For the case of ice SBs, at initial velocities higher than 15~km/s, a complete loss of mass may takes place. 
A faster mass loss of stone SBs compared to iron SBs of the same size is associated with a lower mass of stone SBs due to much lower material density. That is if the amount of absorbed energy from the boundary layer of shock wave  will be the same for equal sizes, but the fraction of the lost mass (relative to the initial one) will be greater for the stone SB compared to the iron one.
Complete loss of mass of the ice SBs at any initial velocities and sizes is explained by their low mass due to low ice density. Another reason is a low specific heat of sublimation of ice. Larger minimum altitudes of trajectory in calculations ($h_{min}$) indicated in the Table caption  for stone and ice SBs were used  to prevent these SBs  from falling.    

Fig.~\ref{fig9a} illustrates the variation of the velocity of SBs along the trajectory  with radii $R=100$, 50 and 25~m. The calculations were carried out  with the values of the drag coefficient $c_d$=0.9 for spherical body.
The obtained results show that the smaller the SB size, the higher the deceleration. There is more notable decrease of the SB velocity with a radius of 25~m in comparison with the radii of 100~m and 50~m. Fig.~\ref{fig9a} contains very important result for further consideration because we come to the conclusion: this dependence lies in the basis of the mechanism of the SB tail formation when fragmentation of the SB occurs -- the smaller the fragment and the lower its kinetic energy, the greater its deceleration and the lower its final velocity   compared to larger fragments.

Fig.~\ref{fig10} demonstrates the changes in trajectories  of the iron SB with $R=25$~m passing through the atmosphere with initial velocities 20 and 15~km/s and the effect of lengthening of the trajectory from 3000 to 5000 km at slower velocity and the minimum trajectory altitude of 11~km. Both SBs exit the atmosphere but at different moments of time.  The effect of lengthening of the SB trajectory is also shown in Fig~\ref{fig7}.   

\section{Application of the through passage model  to the Tunguska phenomenon}

At present, there are over one hundred  hypotheses about the nature of the Tunguska phenomenon, among which, 3--4  versions are predominant theories, e.g.~\citep{Bronshten2000,Farinella2001,Kundt2003,Fesenkov1962,Gladysheva2008}. They include the fall onto the Earth of a small asteroid measuring several dozen metres~\citep{Kundt2003}, consisting of typical asteroid materials, either metal or stone, as well as ice, which is the characteristic of cometary nuclei~\citep{Bronshten2000b,Fesenkov1962,Morrison2019}. The most probable material of the Tunguska SB mentioned in literature is ice. According to the available observational data, there are several variants of the direction and the  trajectory length  of the Tunguska SB~--- from 450 to 600~km, in particular, with a propagation direction from ``south-north'' to ``east-west''. The value of the angle of entry into the atmosphere mentioned in literature is 30--40$^\circ$. The radius of the Tunguska SB was estimated based on the amplitude of the shock wave recorded by the seismic stations and amounted to about 25~m.  
The minimum trajectory altitude of the Tunguska SB approximately corresponded to the point of maximum energy release. 

\begin{figure}
\centering
\begin{tabular}{c}
	\includegraphics[width=5cm]
	{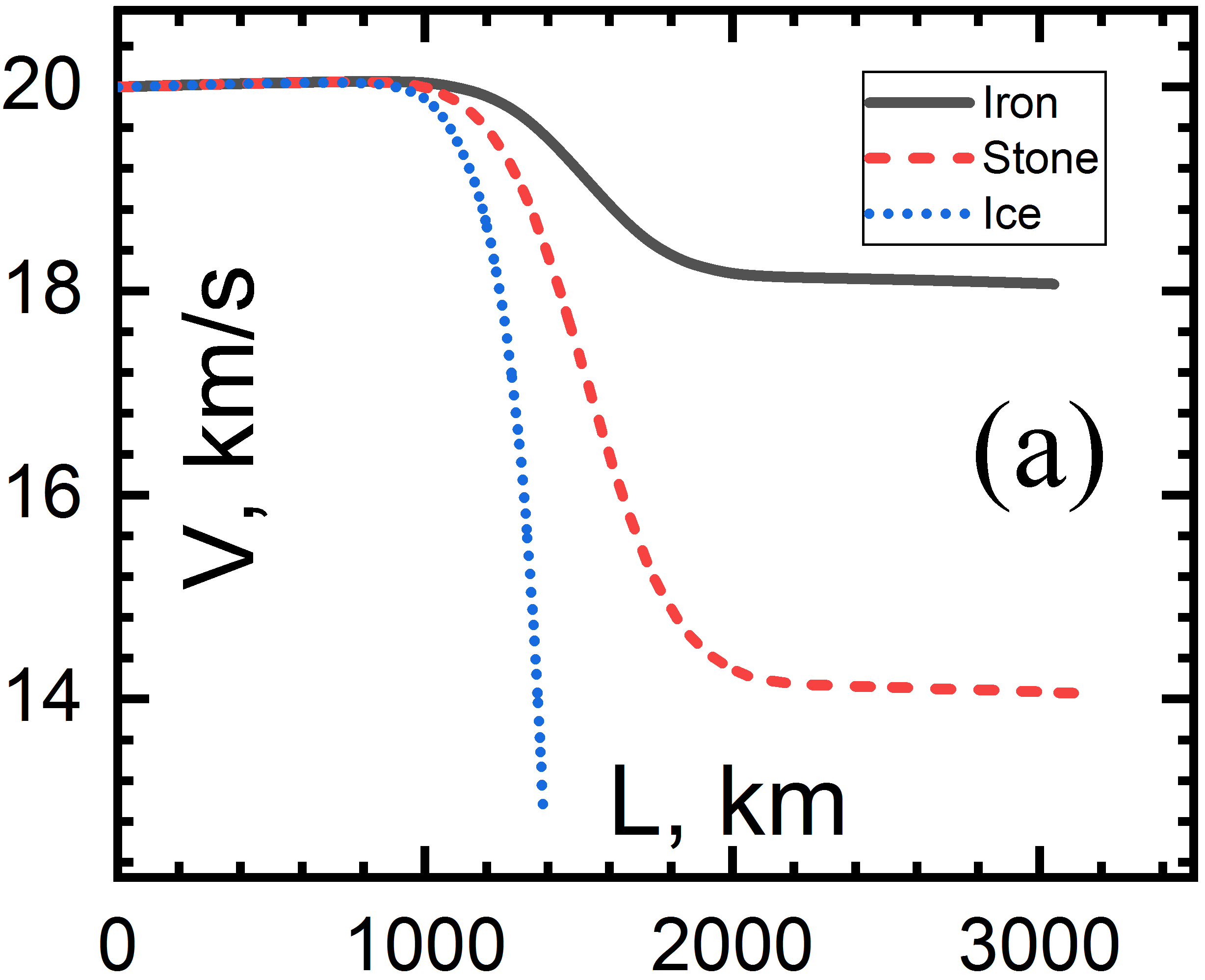}\\
	\includegraphics[width=5cm]
	{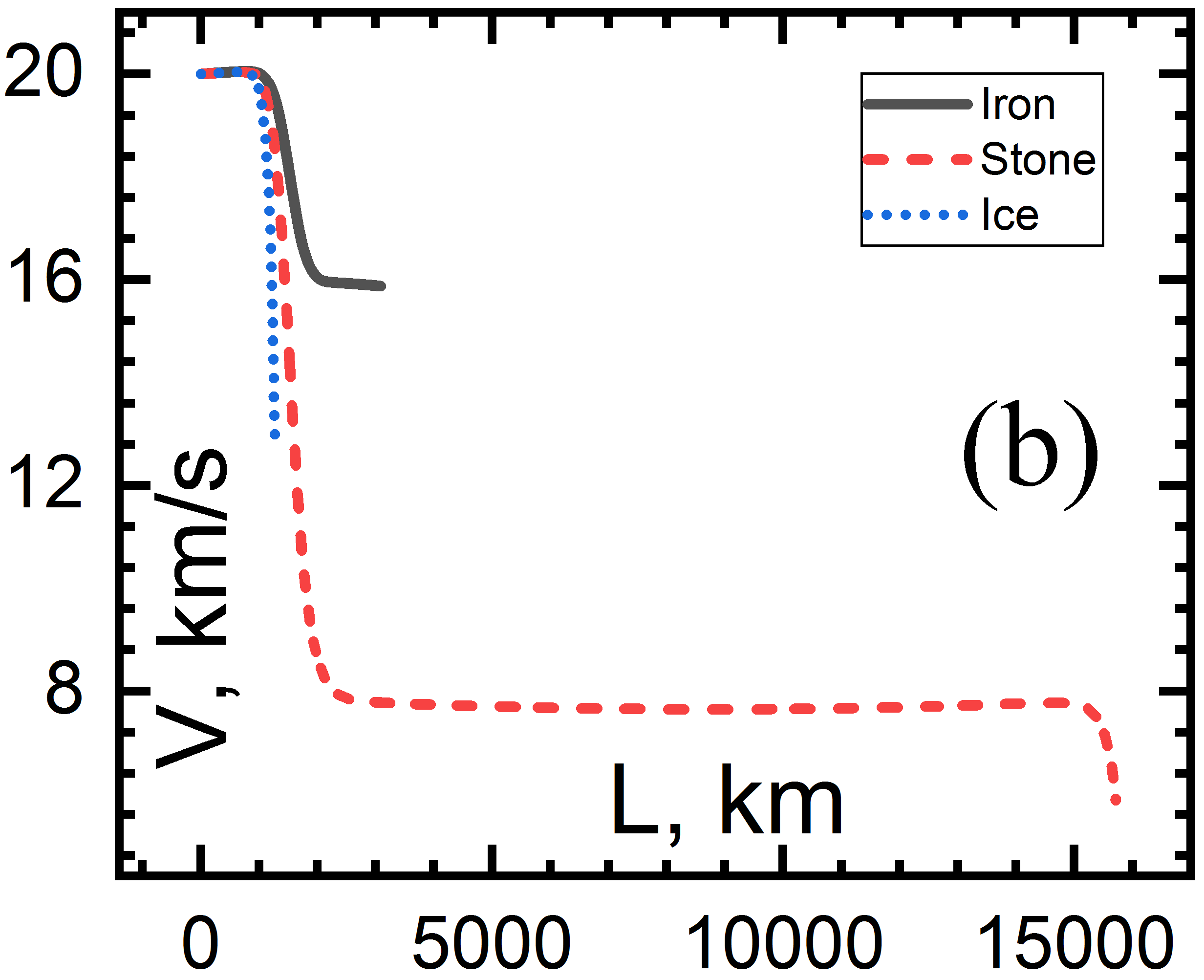}
\end{tabular}
\caption{Comparative variation of the velocity of the iron, stone and ice SB along  the  trajectory with $R=100$~m (a) and $R=50$~m (b). The initial velocity 20~km/s, the minimum trajectory altitude is 11 km. The curves for iron and for stone SB in panel (a) end when the SB exits the atmosphere at the altitude $h=160$~km. The curves for ice SBs end when the mass of the SBs vanishes completely. Long curve for stone SB in panel (b) corresponds to an unusually long trajectory of the stone SB shown in Fig.~\ref{fig12}}
\label{fig11}
\end{figure}

In Fig.~\ref{fig11} we show the results of comparative calculations of the velocity variations  of the iron, stone and ice SBs with radii 100 and 50~m  along the trajectory of through passage across the atmosphere for initial velocity of 20~km/s. Stone SBs lose their velocity faster than iron SBs and ice SBs do not survive passage through the atmosphere. 

\begin{figure}
    \centering
    \includegraphics[width=6.5cm]{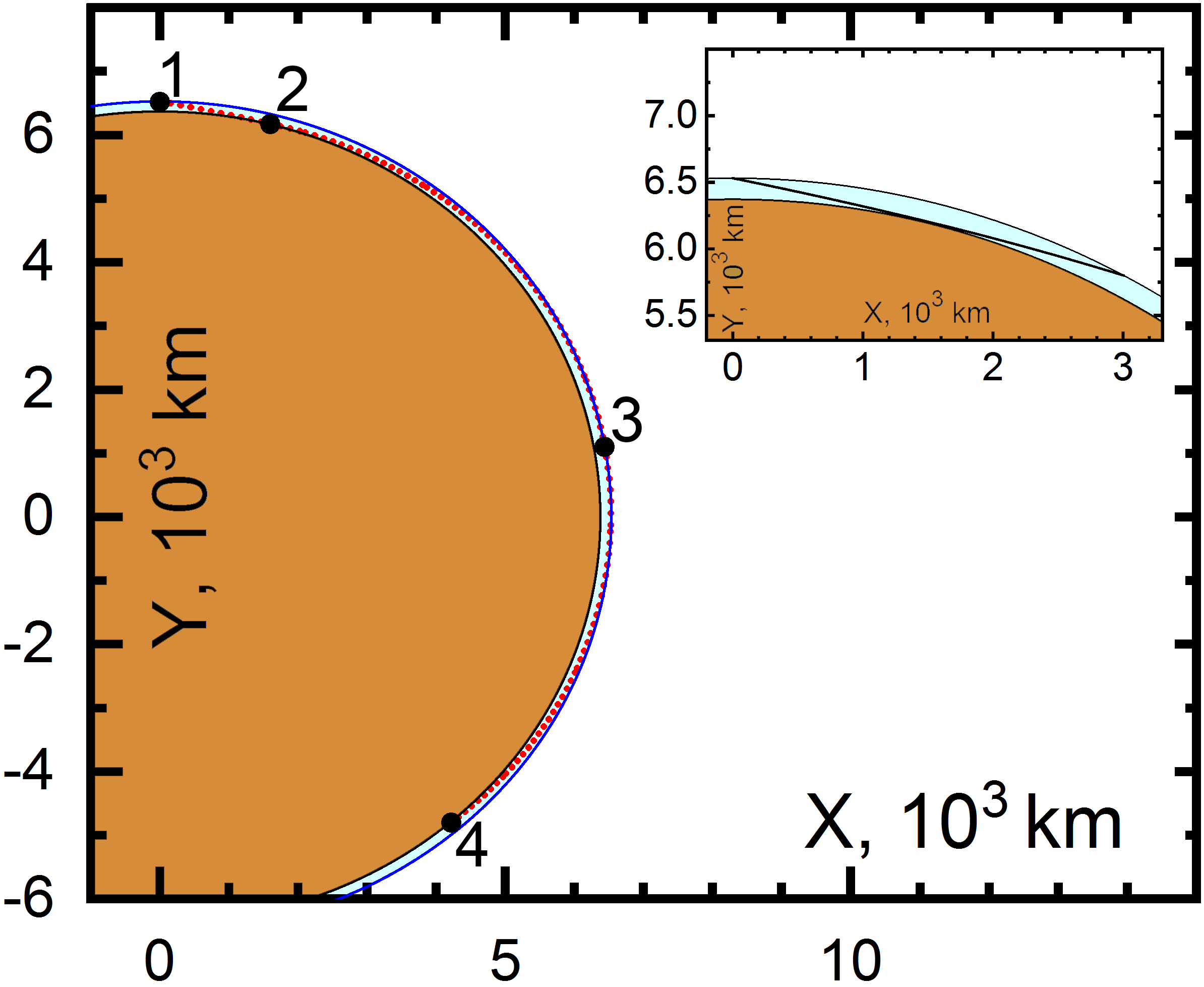}
    \caption{The trajectory of stone SB with radius $R=50$~m. Initial velocity is $V=20$~km/s, minimum altitude is 11~km. 
    In point 1: $V = 20\text{ km/s}$, relative mass $M / M_{in} = 1$, $h = 160\text{ km}$;
    in point 2: $V = 14.2\text{ km/s}$, $M / M_{in} = 0.3$, $h = 11.26\text{ km}$;
    in point 3: $V = 7.6\text{ km/s}$, $M / M_{in} = 0.13$, $h = 159.7\text{ km}$;
    in point 4: $V = 5.9\text{ km/s}$, $M / M_{in} = 0.11$, $h = 11\text{ km}$ 
        (with subsequent fall). The inset shows the comparison of trajectories of stone and iron SBs with 
        equal radii $R=50$~m.}
    \label{fig12}
\end{figure}

\begin{figure}
\centering
	\includegraphics[width=5cm]{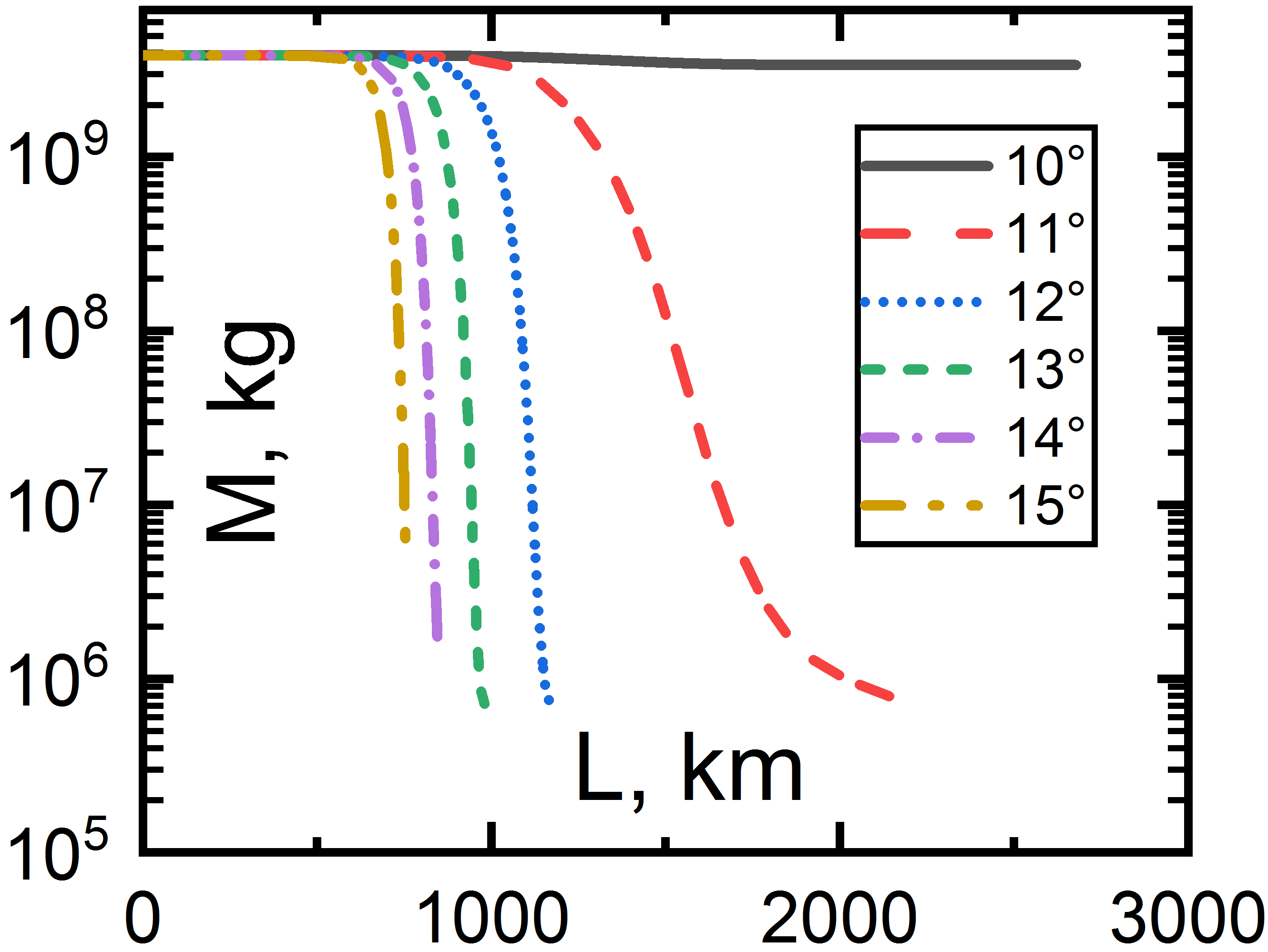}
    \caption{Variation of the ice SB mass along the  trajectory with the initial velocity 20~km/s  and the entry angle in the range $10^\circ\leq\beta\leq15^\circ$ for different minimum altitude  of the trajectory. The initial SB radius is 100~m.}
    \label{fig13}
\end{figure}

In Fig.~\ref{fig12} we demonstrate unusual trajectory of a stone SB with R=50 m compared to iron
one with the same size. At point 1, the SB penetrates the atmosphere at the altitude 160 km, at point 2 it reaches the minimum altitude of 11 km, at point 3 it exits the atmosphere at the altitude 160 km with subsequent re-entry due to a significant decrease of velocity, and at point 4 it is near the point of fall. There is considerable lengthening of trajectory of the stone SB compared to the iron body which passes through the atmosphere with minimum loss of velocity and minimum deflection due to high initial mass  (its trajectory is shown in inset). Although quite improbable, such an SB could manifest itself as a pair of explosive phenomena in the atmosphere separated by thousands kilometres in distance and tens of minutes in time. 

\begin{figure}
\centering
    \begin{tabular}{c}
	\includegraphics[width=5cm]{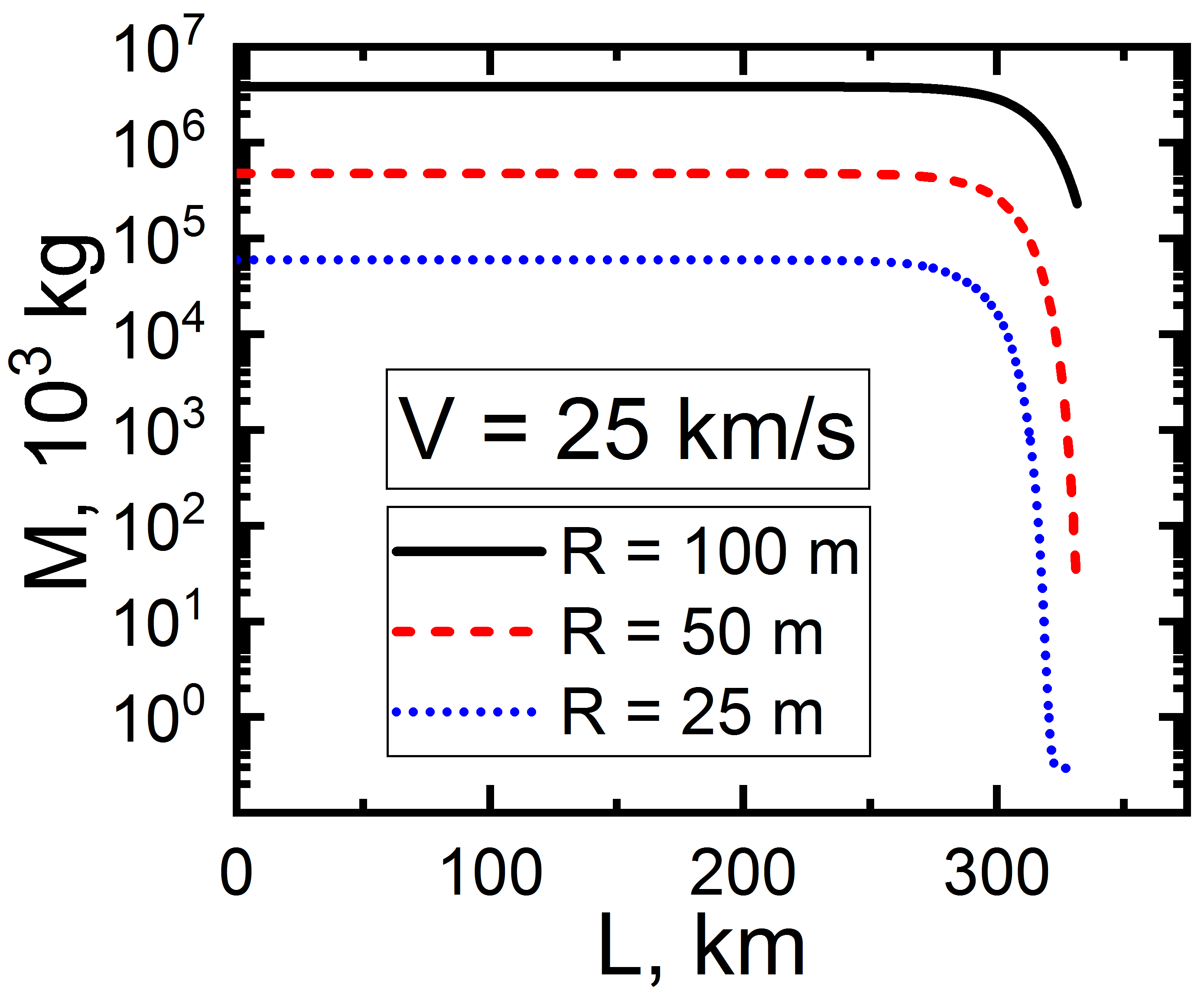}\\
	\includegraphics[width=5cm]{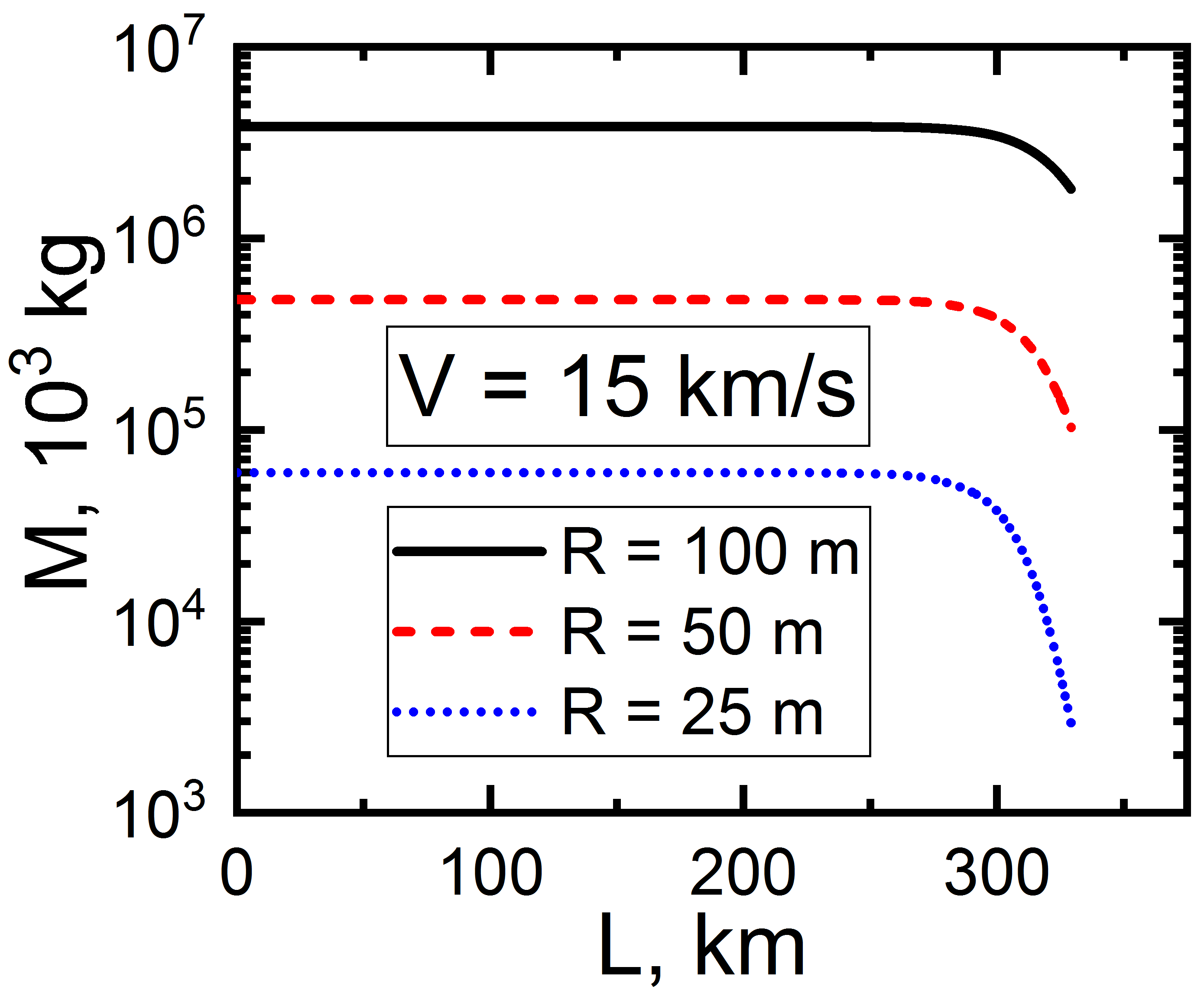}
	\end{tabular}
    \caption{Variation of the ice SB mass along the  trajectory of fall with the initial velocity 25~km/s  and 15~km/s and the entry angle  $30^\circ $. The initial SB radius is 100~m.}
    \label{fig14}
\end{figure}

Fig.~\ref{fig13} shows trajectories of the ice SB with $R=100$~m  at different entry angles and changes in mass. This figure demonstrates dramatic loss of mass at the angles over 11$^\circ$. At the angle 10$^\circ$ the initial mass is preserved due to high altitude~--- over 50~km (Fig.~\ref{fig3}).

Fig.~\ref{fig14} shows   that the residual fractions of the mass of  ice SBs with R=100, 50 and 25 m on the trajectory of fall at  initial velocity of 15~km/s are  49\%, 21.3\% and 4.8\% for radii R=100, 50 and 25~m on the  trajectory of collision with the surface of the Earth. The length until the moment of the collision with the surface of the Earth is about 325~km. At the entry velocity of 25~km/s for radii $R=100$ and 50~m, SBs fall with a preservation of 6\% and 0.00004\% of the initial mass. For radius $R=25$~m an ice SB loses all its mass completely within a trajectory length of about 329~km.

Of course, the fall of SB with preservation of a significant part of the initial mass results in  the formation of a crater with a diameter larger than 1~km~\citep{Stulov1995}. But the fact is that there are no craters near the epicentre and around. The actual length of the trajectory based on the results of visual observations was estimated to be about 450--700~km, which is over 1.5 times longer than the calculated value for the ice SB. Therefore, the hypothesis of the ice origin of the Tunguska SB which enters the atmosphere at the angle $\beta=30$--40$^\circ$, is hardly justified from this point of view. 

Moreover Fig.~\ref{fig13}  shows the decrease of the mass of the ice SB with the initial radius of 100~m along the trajectory at small angles of entry into the atmosphere. As can be seen from the figure, the passage of the ice SB through the atmosphere while preserving the significant fraction of mass is possible only at a minimum altitude above 40~km, which contradicts with the estimated  minimum altitude of about 10--15~km in the Tunguska event.

Our calculations showed that the trajectory length of the ice SB when it passes through the atmosphere at minimum altitude of 15.5~km and small entry angles (less than 15$^\circ$) until the moment of its complete loss of mass even at a radius of 100~m  is two times shorter compared to the case of the iron SB. Thus, the through passage of the ice SB at small entry angles with minimum trajectory altitude 10--15~km is impossible. 

For the ice SB with radius of 25~m, the length of the trajectory to the moment of the total loss of mass is reduced by 4--5 times. In addition, it was shown that a considerable part of the initial mass is preserved by iron and stone SBs with radii of 100, 50 and 25~m at the entry angle of 30$^\circ$. But their fall would be accompanied by the formation of craters with the diameter larger than 1~km and the depth over 200~m.

As the final comments, which can be considered as the plan for future research, we can mention  the following problems. 
     In our work, we did  not deal with the problem of  formation of a shock wave, although  when comparing the Tunguska phenomenon with the Chelyabinsk meteorite with the size of about 10 m and the altitude of maximum energy release of about 30 km  we have no reason to doubt that the body that is 10--20 times larger with  the altitude of maximum energy release of 10-15 km at the velocity of 20 km/s will create a shock wave with a huge amplitude and destructive force, capable of causing tree-fall over an area exceeding 1600 km$^2$.
Experimental modelling of the knock down effect of a shock wave from the source with cylindrical geometry was performed by \citet{Zotkin1966}. The cylindrical source of the shock wave was modelled by a long detonating cord inclined at a certain angle to the plane planted with small sticks, which imitated trees in the Siberian forest. 
It was shown that the shape of the area of fallen sticks was similar to the shape of real tree-fall territory. However, \citet{Zotkin1966} did not model the dependence of the strength of the cylindrical shock wave on the height of its source above the ground. Instead, they added a point explosive at the lower end of their cord to model a presumed spherical component of the shock wave. Because rates of the mass and energy losses of the SB which caused Tunguska event depend strongly on its altitude above the ground (as evident from our Fig.~\ref{fig8a}), sharp increase in energy release close to the minimum altitude reached by the through passing SB can be interpreted as an explosion creating spherical component of the shock wave. Clearly, making a detailed predictions for the patterns of tree-fall in the framework of our hypothesis of a through passed SB as a cause for Tunguska event is an important subject of future research. 

 In solving the main problems in the present work, we confined ourselves to the need to make an upper estimate for calculating the residual mass of space body  using the parameters maximizing the mass loss. 
Finally, in this paper, we did not consider the problem of the mass loss of the space  body due to its fragmentation. This is the subject of future research and the results will be published elsewhere \citep{Khrennikov2020b}.

\section{Summary}

Based on the obtained results we can make the following statements:
\begin{enumerate}[label={\arabic*.}]

    \item The conditions for the possible through passage of a large space body composed of various materials across the Earth's atmosphere with minimal loss of mass and without collision with the surface of the planet are established. It was shown that this corresponds to the entry angles of space body into the atmosphere   $\beta\leq 11.5^\circ$. 
    \item It was shown that the Tunguska space  body could hardly consist of ice, since the length of the trajectory of such a body in the atmosphere before the complete loss of its mass would  be less than the length of its  trajectory   estimated on the basis of observational data.  This statement is valid for estimates performed for the value of  the radiation heat transfer coefficient $c_h$=0.1 as well as making allowance for uncertainties and variations of the values mentioned in the literature.    
    \item The value of the angle of entry into the atmosphere of 30--40$^\circ$ mentioned in the literature for the Tunguska space body looks unrealistic, since it corresponds to the trajectory of a fall of a body with large residual mass and the trajectory length which is 1.5--2 times shorter than the estimated trajectory length based on the observational data. Such a fall would be accompanied by the formation of a large crater, absent near the epicenter and around. 
    \item Probably the most realistic version explaining the Tunguska phenomenon is the through passage of the iron asteroid body as the most resistible to fragmentation across  the Earth's atmosphere at the minimum altitude of 10--15~km with the length of the trajectory in atmosphere of about 3000~km and subsequent exit of this asteroid body into the outer space to the near-solar orbit. This version is supported by the fact that there are no remnants of this body and craters on the surface of the Earth. Within this version we can explain optical effects associated with a strong dustiness of high layers of atmosphere over Europe, which caused bright glow of the night sky.
    
    If we admit the version of the complete loss of mass of SB  after the passage of the epicentre or close to  it, then the evidence of its reality would be the presence of droplets of meteoric iron of the millimetre sizes on the Earth's surface along the trajectory of SB and around. This follows from Fig.~\ref{fig9a}: the smaller the SB size and its mass, the faster it loses a velocity (the amplitude of the shock wave near the epicentre also becomes smaller).  Finally, when the velocity of a diminishing SB reduces to such an extent that its surface temperature approaches 1000~$^\circ$C, the sublimation ceases and the dominant mechanism of the mass loss  consists in blowing off a liquid film from the surface of the body. 
    In this case the SB  becomes the source of a huge amount of droplets, which will be sprayed by SB. However, such micro-formations have not been found despite intensive searches around epicentre and far beyond. The absence of iron droplets around the epicentre is explained by high velocity of the SB  -- over 11.2 km/s when  the surface temperature exceeds several thousands degrees. The dominant mechanism of the mass loss at these temperatures is sublimation of material in the form of  single atoms that can be found on the Earth's surface as iron oxides which do not differ from same widespread terrestrial materials.

\end{enumerate}

\section*{Acknowledgements}

Authors thank A.B. Klyuchantsev for aerodynamic calculations with the software package ANSYS Fluent. The manuscript benefited from many suggestions and comments made in the constructive report by the reviewer, Dr. Darrel Robertson, whom we thank for careful reading of the manuscript. We are grateful to Doug Black of Hamilton, Canada for correcting English in the final version of the manuscript.

%%%%%%%%%%%%%%%%%%%%%%%%%%%%%%%%%%%%%%%%%%%%%%%%%%



\bibliographystyle{mnras}

\bibliography{biblio}

%%%%%%%%%%%%%%%%%%%%%%%%%%%%%%%%%%%%%%%%%%%%%%%%%%

%%%%%%%%%%%%%%%%% APPENDICES %%%%%%%%%%%%%%%%%%%%%

%%%%%%%%%%%%%%%%%%%%%%%%%%%%%%%%%%%%%%%%%%%%%%%%%%

% Don't change these lines
\bsp	% typesetting comment
\label{lastpage}
\end{document}